\documentclass[aps,prd,showpacs,twocolumn,groupedaddress]{revtex4}
\usepackage{graphicx}
\usepackage{amsfonts}
\usepackage{amssymb}

\usepackage{ulem}
\usepackage{color}

\begin{document}

\title{
  Instantaneous interquark potential in generalized Landau gauge
    in SU(3) lattice QCD: a linkage between the Landau and the Coulomb gauges
}
\author{Takumi~Iritani}
\affiliation{Department of Physics, Graduate School of Science, 
  Kyoto University, \\
  Kitashirakawa-oiwake, Sakyo, Kyoto 606-8502, Japan}
\author{Hideo~Suganuma}
\affiliation{Department of Physics, Graduate School of Science, 
   Kyoto University, \\
   Kitashirakawa-oiwake, Sakyo, Kyoto 606-8502, Japan}
\date{\today}

\begin{abstract}

We investigate in detail ``instantaneous interquark potentials'', 
interesting gauge-dependent quantities defined from the spatial correlators 
of the temporal link-variable $U_4$, in generalized Landau gauge 
using SU(3) quenched lattice QCD.
The instantaneous Q$\bar{\rm Q}$ potential has no linear part in the Landau gauge, 
and it is expressed by the Coulomb plus linear potential in the Coulomb gauge,
where the slope is 2-3 times larger than the physical string tension.
Using the generalized Landau gauge, 
we find that the instantaneous potential can be continuously described 
between the Landau and the Coulomb gauges, 
and its linear part rapidly grows in the neighborhood of the Coulomb gauge.
We also investigate the instantaneous 3Q potential in the generalized Landau gauge, 
and obtain similar results to the Q$\bar{\rm Q}$ case.
$T$-length terminated Polyakov-line correlators and their corresponding 
``finite-time potentials'' are also investigated in generalized Landau gauge.
\end{abstract}

\pacs{12.38.Gc,14.70.Dj,12.39.Jh,12.39.Pn}


\maketitle

\section{Introduction}

  Nowadays, quantum chromodynamics (QCD) is established as 
  the fundamental theory of the strong interaction, 
  and perturbative QCD gives a standard framework to describe 
  high-energy reactions of hadrons. 
  QCD is a nonabelian gauge theory constructed from quarks and gluons, 
  and color SU(3) gauge symmetry is one of the guiding principles 
  in formulating QCD \cite{Nambu66,GWP73}.
  For actual perturbative calculations of QCD, 
  the gauge has to be fixed to remove gauge degrees of freedom.
  In the low-energy region, however, QCD exhibits a strong-coupling nature 
  and the resulting nonperturbative QCD is a very difficult and complicated theory.

  As for the gauge fixing, 
  the Landau and the Coulomb gauges have been often used as the typical gauge in QCD, 
  but their physical pictures seem to be rather different 
  for several important arguments in QCD.

  As the typical example, the color confinement, which is an important 
  gauge-invariant QCD phenomenon, can be explained from various viewpoints in various gauges.
  In the Landau gauge, the color confinement is mathematically investigated
  by the Kugo-Ojima criterion, in terms of the BRST charge and the inverse Higgs theorem 
  \cite{KugoOjima}.
  In the Coulomb gauge, the color confinement is argued from the viewpoint 
  of a large instantaneous Coulomb energy \cite{Gribov, Zwanziger98, Greensite03, Greensite04}, 
  and its resulting gluon-chain picture \cite{GreensiteThorn,tHooft}.
  Taking the maximally Abelian (MA) gauge, the quark confinement has been discussed 
  in terms of the dual superconductor picture \cite{NambutHooftMandelstam}.

  Of course, in gauge theories, 
  the physical quantities never depend on the gauge choice.
  However, according to the gauge choice, the physical picture can be changed,
  and the role of the gauge field, which is fundamental field of the gauge
  theory, can be also changed. Then, it is important to link the different gauges,
  and investigate the role of gluons in each gauge.
  In fact, the role and the properties of gluons are expected to be clarified by the
  overview on the structure of gauge dependence.
  For example, it is meaningful to investigate the continuous connection 
  between the Landau and the Coulomb gauges, using the generalized Landau gauge \cite{Bernard90,Iritani10}.

  In this paper, aiming to grasp the gluon properties through
  a continuous view from the Landau gauge to the Coulomb gauge,
  we investigate ``generalized Landau gauge'' (or ``$\lambda$-gauge''),
  and apply it to interesting gluonic correlations, such as 
  instantaneous interquark potentials both for Q$\bar {\rm Q}$ and 3Q systems,  
  in SU(3) lattice QCD at $\beta$=5.8 \cite{Iritani10}.
  Here, the generalized Landau gauge is a natural general gauge to connect 
  the Landau, the Coulomb, and the temporal gauges, 
  by one real parameter $\lambda$.
  In this interpolating gauge, we quantitatively clarify how rapidly the expectation
  values of several operators transition from their Coulomb gauge values to
  the Landau gauge results. 

  The organization of this paper is as follows.
  In Sec.II, we briefly review the properties of the Landau gauge and the Coulomb gauge.
  In Sec.III, we give the formalism of generalized Landau gauge ($\lambda$-gauge).
  In Sec.IV, we formulate the instantaneous potential in lattice QCD.
  In Sec.V, we show the lattice QCD results.
  In Sec.VI, we investigate Polyakov-line correlators and its relation to the potential.
  Sec.VII will be devoted to summary and discussions.

\section{Landau gauge and Coulomb gauge}
  In this section, we briefly review the properties of the Landau gauge and the Coulomb gauge.

  \subsection{Landau gauge}

  The Landau gauge is one of the most popular gauges in QCD, 
  and its gauge fixing is given by 
  \begin{equation}
    \label{eqLandauFix}
    \partial_\mu A_\mu = 0,
  \end{equation}
  where
  $A_\mu(x) \equiv A_\mu^a(x) T^a \in
  \mathfrak{su}(N_c)$ are gluon fields, 
  with $\mathfrak{su}(N_c)$ generator $T^a (a = 1,2,\dots N_c^2-1)$.
  The Landau gauge keeps the Lorentz covariance
  and the global SU($N_c$) color symmetry.
  These symmetries simplify the tensor structure of 
  various quantities in QCD.
  For example, the gluon propagator $D_{\mu\nu}^{ab}(p)$
  is simply expressed as
  \begin{equation}
    D_{\mu\nu}^{ab}(p) = D(p^2) \delta^{ab} \left( 
        g_{\mu\nu} - \frac{p_\mu p_\nu}{p^2} \right),
  \end{equation}
  due to the symmetries and the transverse property.
  Owing to this advanced feature, the Landau gauge 
  is often used both in the Schwinger-Dyson formalism
  \cite{HigashimaMiransky,Alkofer01} and in 
  lattice QCD studies for quarks and gluons \cite{Mandula99,Iritani09}.

  In Euclidean QCD, the Landau gauge has 
  a global definition to minimize the global quantity,
  \begin{equation}
    R_{\rm L} = \int d^4x \mathrm{Tr} \left\{ A_\mu(x) A_\mu(x) \right\}
    = \frac{1}{2}\int d^4x A_\mu^a(x) A_\mu^a(x),
  \end{equation}
  by the gauge transformation.
  This global definition is more strict, 
  and the local form in Eq.(\ref{eqLandauFix}) can be obtained 
  from the minimization of $R_{\rm L}$.
  Since the quantity $R_{\rm L}$ physically means the total amount of gauge-field fluctuations, 
  and therefore the Landau gauge maximally suppresses  
  artificial gauge-field fluctuations
  originated from the gauge degrees of freedom.

  Here, we comment on non-locality of the gauge fields.
  Through the gauge fixing procedure,
  gauge fields have non-locality stemming from the Faddeev-Popov determinant.
  In the Landau gauge, this non-locality of gauge fields
  is Lorentz covariant.

  Using the Landau gauge, or a covariant and globally symmetric gauge,  
  the color confinement has been mathematically investigated in terms of the BRST charge 
  and the inverse Higgs theorem, which is known as the ``Kugo-Ojima criterion'' \cite{KugoOjima}.

  \subsection{Coulomb gauge}

  The Coulomb gauge is also a popular gauge in QCD, 
  and its gauge fixing is given by
  \begin{equation}
    \partial_i A_i = 0.
  \end{equation}
  This condition resembles the Landau gauge
  condition of Eq.(\ref{eqLandauFix}),
  but there are no constraints on $A_0$.
  In the Coulomb gauge, the Lorentz covariance is partially broken, 
  and gauge field components are completely decoupled into two parts, 
  $\vec{A}$ and $A_0$:
  $\vec{A}$ behave as canonical variables and 
  $A_0$ becomes an instantaneous potential.

  Similarly in the Landau gauge, the Coulomb gauge has 
  a global definition to minimize the global quantity
  \begin{equation}
    R_{\rm Coul} \equiv \int d^4 x \mathrm{Tr} 
    \left\{ A_i(x) A_i(x)\right\}
     = \frac{1}{2} \int d^4x A_i^a(x) A_i^a(x)
  \end{equation}
  by the gauge transformation.
  Note here that 
  the Euclidean metric is not necessary for
  the global definition of the Coulomb gauge.
  Note also that there appears no nonlocality in the temporal direction 
  in the Coulomb gauge. Due to this feature, 
  a hadron mass measurement can be safely performed using 
  a spatially-extended quark source in the Coulomb gauge 
  in lattice QCD calculations \cite{CGsource}.

  In the Coulomb gauge, one of the advantages is the compatibility
  with the canonical quantization \cite{ItzyksonZuber}.
  The QCD Hamiltonian is expressed as
  \begin{eqnarray} 
    \label{eqQcdHamiltonian}
    H &=& \frac{1}{2} \int d^3x \left(\vec{E}^{a}\cdot \vec{E}^{a}
        + \vec{B}^{a} \cdot \vec{B}^{a} \right) \nonumber \\
      & &+ \frac{1}{2} \int d^3x d^3y \rho^a(x)K^{ab}(x,y)\rho^b(y), 
  \end{eqnarray}
  where $\rho^a$ is the color charge density, $\vec{E}^a$ and $\vec{B}^a$ are 
  the color electric and magnetic field, respectively.
  Here, $K^{ab}(x,y)$ is the instantaneous Coulomb propagator \cite{Greensite03}
  defined as
  \begin{equation}
    K^{ab}(x,y) = [ M^{-1} (-\nabla^2)M^{-1}]_{xy}^{ab},
  \end{equation}
  with the Faddeev-Popov operator 
  \begin{equation}
    M^{ac} = - \partial^2 \delta^{ac} - \varepsilon^{abc}A_i^b \partial_i.
  \end{equation}

  The confinement picture in the Coulomb gauge focuses 
  on the Coulomb energy including the inverse of $M$.
  Here, the Coulomb energy is the non-local second term of 
  the QCD Hamiltonian (\ref{eqQcdHamiltonian}), 
  and is regarded as the instantaneous potential. 
  Near the Gribov horizon, where the Faddeev-Popov operator $M$ 
  has zero eigenvalues \cite{Gribov}, 
  the Coulomb energy at large quark distance is expected to be largely enhanced 
  and leads to a confining interquark potential, 
  which is called as the ``Gribov-Zwanziger scenario'' \cite{Gribov,Zwanziger98}.

  As Zwanziger showed, the Coulomb energy (instantaneous potential) 
  $V_{\rm Coul}(R)$ in the Coulomb gauge 
  gives an upper bound on the static interquark potential $V_{\rm phys}(R)$
  \cite{Zwanziger03}, i.e.,
  \begin{equation}
    V_{\rm phys}(R) \leq V_{\rm Coul}(R).
  \end{equation}
  This inequality indicates that 
  if the physical interquark potential is confining 
  then the Coulomb energy $V_{\rm Coul}$ is also confining.
  Actually, lattice QCD calculations \cite{Greensite03} 
  show that the Coulomb energy (the instantaneous potential) between a quark and an antiquark
  leads to a linear potential, which characterizes the confinement.
  However, the slope of the instantaneous potential is too large, 
  i.e., $2 \sim 3$ times larger than the physical string tension, 
  and this Coulomb system turns out to be an excited state.

  As for the ground-state of the quark-antiquark system, Thorn and Greensite proposed 
  the ``gluon-chain picture'' in the Coulomb gauge \cite{GreensiteThorn}.
  In fact, to screen the large Coulomb energy between the quark and the antiquark, 
  chain-like gluons are dynamically generated between them.
  This gluon-chain is expected to give the linear potential between quarks. 
  In other words, QCD string can be regarded as a ``chain'' of gluons in the Coulomb gauge.

\section{Generalized Landau gauge}
  In this section, we investigate the ``generalized Landau gauge'', 
  or ``$\lambda$-gauge'', which continuously connects
  the Landau and the Coulomb gauges \cite{Bernard90,Iritani10}.

  \subsection{Definition and Formalism}
  Since the Landau gauge and the Coulomb gauge
  are useful gauges and give different interesting pictures in QCD,
  it is meaningful to show the linkage of these gauges.
  To link these gauges,
  we generalize the gauge fixing condition (\ref{eqLandauFix}) as 
  \begin{equation}
    \label{eqlambdafix}
    \partial_i A_i + \lambda \partial_4 A_4 = 0,
  \end{equation}
  by introducing one real parameter $\lambda$ \cite{Bernard90,Iritani10}.
  The case of $\lambda = 1$ corresponds to the Landau gauge fixing condition,
  the Coulomb gauge is achieved at $\lambda = 0$, 
  and the temporal gauge is also realized for $\lambda \rightarrow \infty$.
  Therefore, we can analyze gauge dependence of various properties
  from the Landau gauge toward the Coulomb gauge
  by varying $\lambda$-parameter from 1 to 0.
  $\lambda$-gauge keeps the global SU($N_c$) color symmetry, but 
  partially breaks the Lorentz symmetry 
  like the Coulomb gauge, except for $\lambda$=1.

  In Euclidean QCD,
  the global definition of $\lambda$-gauge is expressed by 
  the minimization of
  \begin{equation}
    \label{eqgloballambda}
    R^{\lambda} \equiv \int d^4x \left[
    \mathrm{Tr}\left\{A_i(x) A_i(x)\right\}
    + \lambda \mathrm{Tr}\left\{A_4(x) A_4(x)\right\}
    \right]
  \end{equation}
  by the gauge transformation, and 
  this minimization actually leads to Eq.(\ref{eqlambdafix}).
  Here, the $\lambda$-parameter controls 
  the ratio of the gauge-field fluctuations of
  $\vec{A}$ and $A_4$.

  Lattice QCD is formulated on the discretized Euclidean space-time, 
  and the theory is described with the link-variable 
  $U_\mu(x) \equiv  e^{iagA_\mu(x)}\in {\rm SU}(N_c)$, 
  with the lattice spacing $a$ and the gauge coupling constant $g$,
  instead of gauge fields $A_\mu(x) \in \mathfrak{su}(N_c)$.
  $\lambda$-gauge fixing condition is expressed
  in terms of the link-variable as the maximization of a quantity 
  \begin{equation}
    \label{eqRlambda}
    R_{\rm latt}^\lambda[U] \equiv \sum_x \left\{ \sum_i \mathrm{Re}
    \mathrm{Tr} U_i(x) + \lambda \mathrm{Re} \mathrm{Tr} U_4(x) \right\}
  \end{equation}
  by the gauge transformation of the link-variables, 
  \begin{equation}
    U_\mu(x) \rightarrow \Omega(x) U_\mu(x) \Omega^\dagger(x+\hat{\mu}),
  \end{equation}
  with the gauge function $\Omega(x) \in \mathrm{SU}(3)$.
  In the continuum limit of $a \rightarrow 0$, this
  condition results in the minimization of $R^{\lambda}$ in Eq.(\ref{eqgloballambda}), and 
  satisfies the local $\lambda$-gauge fixing condition of Eq.(\ref{eqlambdafix}).

  Note here that the gluon-field fluctuation is strongly suppressed 
  in the generalized Landau gauge with $\lambda \ne 0$, so that 
  one can use the expansion of the link-variable  
  $U_\mu(x) \equiv e^{iagA_\mu(x)} \simeq 
  1 + iagA_\mu(x) + \mathcal{O}(a^2)$ for small lattice spacing $a$, 
  and the gluon field $A_\mu(x)$ 
  can be defined by
  \begin{equation}
    \label{eqgluon}
    A_\mu(x) \equiv 
    \frac{1}{2iag}\left[U_\mu(x)- U_\mu^\dagger(x)\right]_{\rm traceless} 
    \in \mathfrak{su}(N_c)
  \end{equation}
  without suffering from large gluon fluctuations stemming from the gauge degrees of freedom.
  With this gluon field $A_\mu(x)$, the local $\lambda$-gauge condition is expressed as 
  \begin{equation}
    \sum_{i=1}^3 [ A_i(x) - A_i(x-\hat{i})] + \lambda [A_4(x) - A_4(x-\hat{4})]=0,
  \end{equation}
  which directly corresponds to Eq.(\ref{eqlambdafix}) in the continuum formalism.

  \subsection{Convergence into Coulomb gauge for $\lambda \rightarrow 0$}
  In this subsection, we briefly discuss the correspondence between the Coulomb gauge
  and the $\lambda \rightarrow 0$ limit of generalized Landau gauge.

  We comment on the residual gauge degrees of
  freedom in the Coulomb gauge.
  In lattice QCD,
  the Coulomb gauge fixing condition is expressed by
  the maximization of the quantity
  \begin{equation}
    R_{\rm Coul}[U] \equiv \sum_{\vec{x},t} \sum_{i=1}^3
      \mathrm{Re} \ \mathrm{Tr} \ U_i(\vec{x},t)
  \end{equation}
  by the gauge transformation.

  Now, we consider the spatially-global gauge transformation as
  \begin{eqnarray}
    U_i(\vec{x},t) &\rightarrow & \Omega(t) U_i(\vec{x},t) \Omega^\dagger(t), \\
      U_4(\vec{x},t) &\rightarrow & \Omega(t) U_4(\vec{x},t) \Omega^\dagger(t+1),
  \end{eqnarray}
  with the gauge function $\Omega(t) \in {\rm SU}(N_c)$.
  $R_{\rm Coul}[U]$ is invariant under this transformation,
  \begin{eqnarray}
    R_{\rm Coul}[U] &\rightarrow& 
      \sum_{\vec{x},t} \sum_{i=1}^3 \mathrm{Re} \ \mathrm{Tr} \
      \Omega(t) U_i(\vec{x},t) \Omega^\dagger(t) \nonumber \\
    &=& \sum_{\vec{x},t} \sum_{i=1}^3 \mathrm{Re} \ \mathrm{Tr} \ U_i(\vec{x},t).
  \end{eqnarray}
  Therefore, the Coulomb gauge has the corresponding residual gauge degrees of freedom.

  Under this residual symmetry, however,  $\mathrm{Tr} \ U_4$ is gauge-variant as 
  \begin{eqnarray}
    \mathrm{Tr} \ U_4(\vec{x},t) &\rightarrow&
    \mathrm{Tr} \ \Omega(t) U_4(\vec{x},t) \Omega^\dagger(t+1) 
  \end{eqnarray}
  so that the expectation value $\langle \mathrm{Tr} \ U_4 \rangle$ is to be zero
  in the Coulomb gauge.

  In the generalized Landau gauge with non-zero $\lambda$-parameter,
  this residual symmetry does not exist,
  and hence $\langle \mathrm{Tr} \ U_4 \rangle$ has a finite value,
  as will be discussed in Sec.VI.

  We here investigate the convergence of the 
  generalized Landau gauge into the Coulomb gauge in the limit of $\lambda \rightarrow 0$.
  To check the convergence, we evaluate the quantity,
  \begin{eqnarray} 
    \langle \left(\partial_i A_i^a\right)^2 \rangle   &\equiv &
    \frac{1}{(N_c^2-1){N_{\rm site}}} \nonumber \\
    &\times &
    \sum_{x=1}^{N_{\rm site}} \sum_{a=1}^{N_c^2-1} 
    \Big\{ \sum_{i=1}^3 \left[ A_i^a(x) - A_i^a(x-\hat{i})\right]\Big\}^2.~~~~~~~
  \end{eqnarray}
  In the Coulomb gauge, $\langle (\partial_i A_i^a)^2 \rangle$ is equal to zero.

\begin{figure}[h]
    \centering
    \includegraphics[width=8cm,clip]{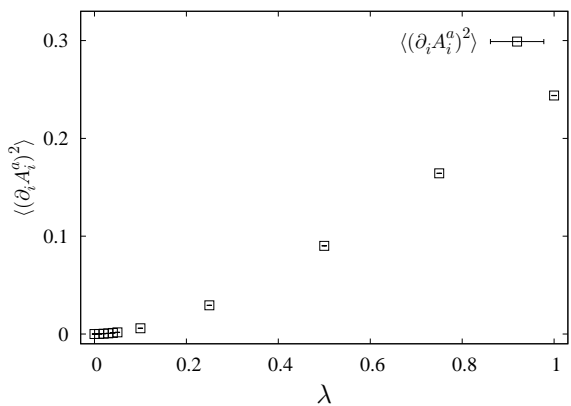}
    \caption{\label{figConvergence}
      The lattice QCD result of 
      $\langle (\partial_i A_i^a)^2 \rangle$ in the generalized Landau gauge.
        As $\lambda \rightarrow 0$, this quantity
        goes to zero monotonically.
      }
  \end{figure}

  Figure \ref{figConvergence} shows $\langle (\partial_i A_i^a)^2 \rangle$ 
  in lattice QCD calculation with $\beta=5.8$ and $16^4$, 
  and we find that $\langle (\partial_i A_i^a)^2 \rangle$ is 
  monotonically decreasing toward zero by varying $\lambda$ from 1 to 0. 
  This result supports that the generalized Landau gauge approaches 
  to the Coulomb gauge in the $\lambda \rightarrow 0$ limit.

\section{Polyakov-line correlators and instantaneous potential}
  In this section, we formulate Polyakov-line correlators
  and instantaneous potential in lattice QCD.

  First, we consider the Euclidean continuum theory to make clear the physical
  interpretation of the terminated Polyakov-line correlator.
  Considering the source field $J_\mu(x)$ which couples
  to the gauge field $A_\mu(x)$, the generating functional is given by
  \begin{equation}
    Z[J] \equiv \langle \ e^{i\int d^4x \mathcal{L}_{\rm int}(J)} \rangle
    = \langle \exp\{ ig \int d^4x A_\mu^a(x) J^\mu_a(x)\} \rangle.
  \end{equation}
  We consider a closed-loop current such as the Wilson loop $W(R,T)$, 
  which is constructed in a gauge-invariant manner.
  From the relation $Z[J] \propto e^{-E_JT}$, 
  the ground-state energy between static quark and antiquark pair is expressed as
  \begin{equation}
    V_{\rm phys}(R) = - \lim_{T \rightarrow \infty} \frac{1}{T} \ln \langle W(R,T) \rangle.
  \end{equation}

  Next, we consider the color current $J_\mu(x)$ of 
  a quark located at $\vec{x} = \vec{a}$ and an antiquark at $\vec{x} = \vec{b}$.
  For the case that these sources are generated at $t = 0$ and annihilated at $t = T$, 
  the color current $J_\mu(x)$ is expressed as  
  \begin{equation}
    J_\mu(x) = \delta_{\mu 4}\left[ \delta(\vec{x}-\vec{a}) - \delta(\vec{x}-\vec{b})\right]
    \theta(T-t)\theta(t),
  \end{equation}
  in the generalized Landau gauge.
  In this case, the current $J_\mu$ is not conserved, and it breaks the gauge invariance.   
  In a similar manner to the Wilson loop $W(R,T)$,
  we define the ``energy'' of two sources
  in the presence of $J_\mu(x)$ as
  \begin{equation}
    \label{eqterminatedV}
    V(R,T) = - \frac{1}{T} \ln \langle \mathrm{Tr} [L(\vec{a},T) L^\dagger(\vec{b},T)] \rangle
  \end{equation}
  with $R = |\vec{a}-\vec{b}|$,
  using the terminated Polyakov-line
  \begin{equation}
    \label{eqterminatedPL}
    L(\vec{x},T) \equiv P \exp\left(ig\int_0^T dx_4 A_4(\vec{x},x_4)\right),
  \end{equation}
  with $P$ being the path-ordered product.
  As a caution, for $\lambda \ne$ 0, $V(R,T)$ is gauge-dependent and
  does not mean the energy of a physical state, due to the temporal nonlocality of
  the Faddeev-Popov determinant.

  Next,
  we consider $N_x \times N_y \times N_z \times N_t$ lattice
  with the lattice spacing $a$.
  Using temporal link-variables, 
  the terminated Polyakov-line with length $T$ is defined as
  \begin{equation}
    L(\vec{x},T) = U_4(\vec{x},a) U_4(\vec{x},2a)\cdots U_4(\vec{x},T).
  \end{equation}
  The terminated Polyakov-line is generally gauge variant, 
  and its expectation value depends on the choice of gauge.
  Only for $T = N_t$, the trace of the Polyakov-line coincides the Polyakov loop, 
  which is gauge invariant. 
  In the Coulomb gauge, the expectation value of the terminated Polyakov-line 
  is zero due to the remnant symmetry, as was discussed in Sec.III B.

  In this paper, we consider the Polyakov-line correlator 
  in generalized Landau gauge denoted by 
  \begin{equation}
    G_\lambda(R,T)
      = \langle \mathrm{Tr} [L^\dagger(\vec{x},T)L(\vec{y},T)]\rangle
  \end{equation}
  with $R = |\vec{x} -\vec{y}|$.
  From this correlator, we define ``finite-time potential'', 
  \begin{equation}
    V_\lambda(R,T) \equiv - \frac{1}{T} \ln G_\lambda(R,T).
  \label{eqFTPot}
  \end{equation}
  Here, for the simple expression, 
  a normalization factor 1/3 is dropped off for $G_\lambda(R,T)$, 
  since it only gives a constant term in the potential $V_\lambda(R,T)$.

  Particularly for $T = a$, we call 
  \begin{equation}
    \label{eqInstantaneousPotential}
    V_\lambda(R) \equiv V_\lambda(R,a) 
    = - \frac{1}{a} \ln \langle \mathrm{Tr}
    \big[ U_4^\dagger(\vec{x},a) U_4(\vec{y},a)\big]\rangle
  \end{equation}
  as ``instantaneous potential''.

  Also, in the generalized Landau gauge, 
  we define ``finite-time 3Q potential'' as 
  \begin{equation}
    V^{\rm 3Q}_\lambda(\vec{x}_1,\vec{x}_2,\vec{x}_3, T)
      \equiv - \frac{1}{T} \ln G^{\rm 3Q}_\lambda(\vec{x}_1,\vec{x}_2,\vec{x}_3),
  \end{equation}
  using the three Polyakov-line correlator of
  \begin{eqnarray}
    &&G^{\rm 3Q}_\lambda(\vec{x}_1,\vec{x}_2,\vec{x}_3,T) \nonumber \\
      &&\equiv \langle
      \varepsilon_{abc}\varepsilon_{a'b'c'}
    L^{aa'}(\vec{x}_1,T)L^{bb'}(\vec{x}_2,T)L^{cc'}(\vec{x}_3,T)\rangle,
  \end{eqnarray}
  which is formulated in a similar manner for
  3Q Wilson loop $W_{\rm 3Q}$ \cite{STI,TS0304}.
  For $G^{\rm 3Q}_\lambda$, a normalization factor $1/3!$ is dropped off, 
  since it only gives a constant term in the potential $V^{\rm 3Q}_\lambda$.
  For $T = a$, we call
  \begin{equation}
    V^{\rm 3Q}_\lambda(\vec{x}_1,\vec{x}_2,\vec{x}_3) \equiv
    V^{\rm 3Q}_\lambda(\vec{x}_1,\vec{x}_2,\vec{x}_3,a)
  \end{equation}
  as ``instantaneous 3Q potential''.

  Here, these quantities depend on $\lambda$-parameter.
  In the Coulomb gauge, the instantaneous potential $V_{\lambda=0}(R)$ 
  (or the Coulomb energy) gives a linear potential, but 
  its slope is about $2 \sim 3$ times larger than
  the physical string tension \cite{Greensite03}.
  In the Landau gauge, 
  the instantaneous potential $V_{\lambda=1}(R)$ has no linear part \cite{Nakamura06},
  which is also expected from the exponential reduction of the gluon propagator 
  \cite{Mandula99,Iritani09}
  and the Lorentz symmetry.
  In Sec.VII, we will discuss the relation between the gluon propagator
  and the instantaneous potential in the Landau gauge.

\section{Lattice QCD results}

  We perform SU(3) lattice QCD Monte Carlo calculations
  at the quenched level.
  We use the standard plaquette action \cite{Rothe} with the lattice parameter 
  $\beta \equiv \frac{2N_c}{g^2}=5.8$ on a $16^4$-size lattice.
  The lattice spacing $a$ is 0.152 fm,
  which is determined so as to reproduce the string tension as
  $\sqrt{\sigma} = 427$ MeV \cite{STI}.

  We use the gauge configurations, which are picked up every
  1000 sweeps after a thermalization of 20000 sweeps.
  After the generation of gauge configurations, 
  we perform gauge fixing by maximizing $R^{\lambda}_{\rm latt}[U]$.
  In this paper, we use the Landau gauge ($\lambda$=1),
  the Coulomb gauge ($\lambda$=0), and 
  their intermediate gauges with $\lambda = 0.75, 0.50, 0.25, 0.10,
   0.05, 0.04, 0.03, 0.02, 0.01$.
  We investigate in detail the region near the Coulomb gauge ($\lambda$=0), 
  since the behavior of the instantaneous potential 
  largely changes for $\lambda \sim $0, as will be shown later.
  The number of gauge configurations is 50 for each $\lambda$.
  We adopt the jackknife method to estimate the statistical error.

  Here, we comment on $\lambda$-gauge fixing convergence.
  We fix the gauge by maximizing the quantity $R^\lambda_{\rm latt}[U]$ 
  in Eq.(\ref{eqRlambda}), which corresponds to 
  $\partial_i A_i + \lambda \partial_4 A_4 = 0$.
  Therefore, to check the convergence of gauge fixing, we evaluate $\epsilon_\lambda$ 
  defined by
  \begin{eqnarray} 
    \epsilon_\lambda &\equiv& 
    \langle \left(\partial_i A_i^a + \lambda \partial_4 A_4^a \right)^2 \rangle \nonumber \\
      &\equiv &
      \frac{1}{(N_c^2-1){N_{\rm site}}}\sum_{x=1}^{N_{\rm site}} \sum_{a=1}^{N_c^2-1}
      \Big\{ \sum_{i=1}^3 \left[ A_i^a(x) - A_i^a(x-\hat{i})\right] \nonumber \\
        & + & \lambda \left[ A_4^a(x) - A_4^a(x-\hat{4})\right] \Big\}^2,
    \label{eqConvergence}
  \end{eqnarray}
  with the gluon field $A_\mu(x)=A_\mu^a(x)T^a$ given in Eq.(\ref{eqgluon}).
  We iterate the gauge transformation 
  to satisfy $\epsilon_\lambda < {10}^{-12}$ finally.
  As for the instantaneous potential $V_\lambda(R)$,
  this convergence condition is very strict.
  Actually, we can obtain stable lattice data of $V_\lambda(R)$
  even with $\epsilon_\lambda < {10}^{-4}$.

  Also, we comment that the calculation cost of the gauge fixing is rapidly increasing
  as $\lambda$ approaches to zero, while the Coulomb gauge ($\lambda=0$)
  itself can be easily achieved.
  Considering this critical slowing down of the gauge fixing \cite{Bernard90,Cucchieri07},
  we adopt a relatively small-size lattice of $16^4$ with $\beta = 5.8$, 
  although its physical volume of about $(2.4 {\rm fm})^4$ 
  is large enough to extract the relevant region for the interquark potential.

 \subsection{``Instantaneous Q$\bar {\bf Q}$ inter-quark potential'' 
   in generalized Landau gauge}

\begin{figure}
    \centering
    \includegraphics[width=8cm,clip]{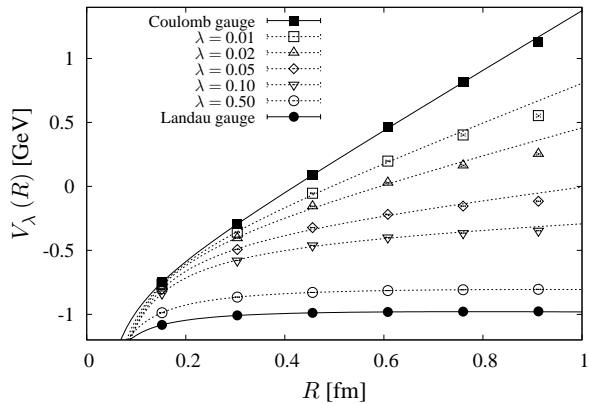}
    \caption{\label{figInstaPot}
      ``Instantaneous potential'' $V_\lambda(R)$ in generalized Landau gauge
        for typical values of $\lambda$.  The symbols denote lattice QCD results, and
        the curves fit-results using Coulomb plus linear form, 
        $V_\lambda(R) = -A_\lambda/R+\sigma_\lambda R +C_\lambda$,
        in the region of $R \lesssim 0.8$ fm.  
        For $\lambda = 1 \sim 0.1$, the potential has almost no linear part.
        For $\lambda \lesssim 0.1$, the linear potential grows rapidly, and 
        $\sigma_\lambda \simeq 2.6\sigma_{\rm phys}$ at $\lambda = 0$.
     }
  \end{figure}

  We investigate the instantaneous potential 
  $V_\lambda(R)$ defined by Eq.(\ref{eqInstantaneousPotential}) 
  in generalized Landau gauge.
  Figure \ref{figInstaPot} shows lattice QCD results of $V_\lambda(R)$
  for typical values of $\lambda$.
  In this figure, the statistic error is small and the error bars are hidden in the symbols.

  In the Coulomb gauge ($\lambda = 0$), the instantaneous potential 
  shows linear behavior, while there is no linear part at all 
  in the Landau gauge ($\lambda = 1$).
  Thus, there is a large gap between these gauges 
  in terms of the instantaneous potential.
  In our framework, however, these two gauges are connected continuously.

  By varying the $\lambda$-parameter from 1 to 0 in the generalized Landau gauge, 
  we find that the instantaneous potential $V_\lambda(R)$ changes continuously, and  
  the infrared slope of the potential $V_\lambda(R)$ at $R \simeq 0.8{\rm fm}$ 
  grows monotonically,
  from the Landau gauge to the Coulomb gauge, as shown in Fig.~\ref{figInstaPot}.
  Note here that 
  the growing of the infrared slope of $V_\lambda(R)$ is quite rapid
  for $\lambda \lesssim 0.1$ near the Coulomb gauge,
  while the infrared slope is rather small and almost unchanged for $\lambda = 0.1 \sim 1$.

  To analyze the instantaneous potential $V_\lambda(R)$ quantitatively,
  we fit the lattice QCD results using 
  the Coulomb plus linear Ansatz as
  \begin{equation}
    V_\lambda(R) = - \frac{A_\lambda}{R} + \sigma_\lambda R + C_\lambda,
    \label{eqCoulplusLin}
  \end{equation}
  where $A_\lambda$ is the Coulomb coefficient and $C_\lambda$  a constant.
  Here, $\sigma_\lambda$ is the infrared slope of the potential, 
  which we call as ``instantaneous string tension''.
  Besides the Coulomb plus linear Ansatz, 
  we try several candidates of the functional form,
  $-A/R + \sigma(1-e^{-\varepsilon R})/\varepsilon$, $-A\exp(-mR)/R$,
  $-A/R + \sigma R^d$, and $-A/R^d$ apart from an irrelevant constant, 
  but they are less workable.
  The curves in Fig. \ref{figInstaPot} are the best-fit results using
  Eq.(\ref{eqCoulplusLin}). 
  The Coulomb plus linear Ansatz works well 
  at least for $R \lesssim 0.8$fm, 
  which is relevant region for hadron physics.
  We note that the Yukawa form of $-Ae^{-mR}/R$ 
  also works well near the Landau gauge, which will be discussed   
  in relation to the gluon-propagator behavior in Sec.VI-C.

  We summarize the best-fit parameters and the fit-range in Table \ref{tabFitResult}.
  While the Coulomb coefficient $A_\lambda$ has a relatively weak $\lambda$-dependence, 
  the instantaneous string tension $\sigma_\lambda$ shows a strong $\lambda$-dependence near 
  the Coulomb gauge, i.e., $\lambda \lesssim 0.1$.

  We comment on the asymptotic behavior of the instantaneous potential.
  In the deep-IR limit of $R \rightarrow \infty$, $V_\lambda(R)$ 
  goes to a saturated value, so that the asymptotic value of 
  the instantaneous string tension goes to zero, except for $\lambda = 0$.
  This behavior is due to $\langle U_4\rangle \ne 0$ for the temporal 
  link-variable $U_4$ in the generalized Landau gauge for $\lambda \ne 0$,
  as will be discussed in Sec.VI.

  \begin{table}
    \begin{center}
    \caption{\label{tabFitResult}
      The best-fit parameters on the instantaneous potential using
      $V_\lambda(R) = - A_\lambda/R + \sigma_\lambda R + C_\lambda$,
      and the ratio of the slope $\sigma_\lambda$ to the physical string tension 
      $\sigma_{\rm phys}$.
      The standard parameters of the physical interquark potential 
      are $A_{\rm phys} \simeq 0.27$ and $\sigma_{\rm phys} \simeq 0.89$GeV/fm \cite{STI}.
      The string tension $\sigma_\lambda$ is rather small for $\lambda = 0.1 \sim 1$.}
    \begin{tabular}{cccccc}
      \hline
      \hline 
      $\lambda$ & fit-range [fm] & $A_\lambda$ & $\sigma_\lambda$ [GeV/fm] 
        & $C_\lambda$ [GeV] & $\sigma_\lambda/\sigma_{\rm phys}$ \\
      \hline
      0.00 & 0.1-1.0 & 0.167(11) & 2.283(35) & -0.881(20) & 2.57(4) \\
      0.01 & 0.1-0.8 & 0.287(27) & 1.476(78) & -0.617(46) & 1.66(9) \\
      0.02 & 0.1-0.8 & 0.346(32) & 1.005(90) & -0.481(54) & 1.13(10) \\
      0.03 & 0.1-0.8 & 0.372(32) & 0.728(86) & -0.416(53) & 0.82(10) \\
      0.04 & 0.1-0.8 & 0.382(30) & 0.557(79) & -0.386(50) & 0.63(9) \\
      0.05 & 0.1-0.8 & 0.386(29) & 0.441(73) & -0.372(47) & 0.50(8) \\
      0.10 & 0.1-0.8 & 0.365(20) & 0.169(46) & -0.390(31) & 0.19(5) \\
      0.25 & 0.1-0.8 & 0.281(6) & -0.005(13) & -0.544(9) & -0.01(1) \\
      0.50 & 0.1-0.8 & 0.198(0) & -0.042(1) & -0.724(1) & -0.05(0) \\
      0.75 & 0.1-0.8 & 0.152(1) & -0.043(3) & -0.839(2) & -0.05(0) \\
      1.00 & 0.1-0.8 & 0.123(2) & -0.040(3) & -0.917(3) & -0.04(0) \\
      \hline 
      \hline
    \end{tabular}
    \end{center}
  \end{table}

\begin{figure}
    \centering
    \includegraphics[width=8cm,clip]{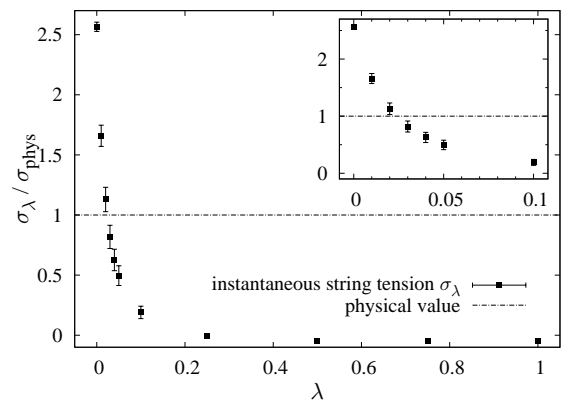}
    \caption{ \label{figStringTension}
      Instantaneous string tension $\sigma_\lambda$, the slope of 
      the linear part in the instantaneous potential $V_\lambda(R)$ 
      in the generalized Landau gauge.
      The right upper figure is a close-up near the Coulomb gauge ($\lambda$=0).
      For $0 \le \lambda \lesssim 0.1$,  
      $\sigma_\lambda$ changes rapidly from $2.6 \sigma_{\rm phys}$ to 0,
      while $\sigma_\lambda$ is rather small for $0.1 \lesssim \lambda \le 1$.
    }
  \end{figure}

  Now, we focus on the $\lambda$-dependence of 
  instantaneous string tension $\sigma_\lambda$ in Fig.~\ref{figStringTension}.
  For $0.1 \lesssim \lambda \le 1$, including the Landau gauge ($\lambda$=1), 
  $\sigma_\lambda$ is almost zero, so that this region can be regarded as ``Landau-like.''
  For $\lambda \lesssim 0.1$, $V_\lambda(R)$ is drastically changed near the Coulomb gauge, 
  and $\sigma_\lambda$ grows rapidly in this small region.
  Finally, in the Coulomb gauge ($\lambda$=0), one finds 
  $\sigma_\lambda \simeq 2.6\sigma_{\rm phys}$, with $\sigma_{\rm phys} \simeq 0.89$GeV/fm.

  Note that the instantaneous string tension 
  $\sigma_\lambda$ continuously changes from $0$ to  
  $(2 \sim 3)\sigma_{\rm phys}$, 
  according to the change from the Landau gauge to the Coulomb gauge, 
  and therefore there exists some specific $\lambda$-parameter of $\lambda_C \in [0,1]$
  where the slope of the instantaneous potential $V_\lambda(R)$
  coincides with the physical string tension $\sigma_{\rm phys}$ for $R \lesssim 0.8$fm. 
  Since the instantaneous potential generally depends 
  on the lattice parameter $\beta$, i.e., the lattice spacing $a$ 
  \cite{Greensite03,Iritani10}, the value of $\lambda_C$ is 
  $\beta$-dependent, although its dependence would be rather weak, 
  as will be discussed in Sec.VI.
  However, from the continuity between 
  the overconfining potential in the Coulomb gauge
  and the saturated potential in the Landau gauge,
  there must exist $\lambda_C \in [0,1]$ 
  where the instantaneous string tension $\sigma_\lambda$
  coincides with $\sigma_{\rm phys}$ for each lattice spacing.

  On the relation to the confinement, which is a gauge independent phenomenon, 
  the role of gluons generally depends on the choice of gauges,
  and the physical picture of the confinement would be changed according to gauges.
  For example, in the Coulomb gauge, 
  the instantaneous Coulomb energy gives an overconfining potential, 
  and the ground-state of the quark-antiquark system is described as the gluon-chain state
  \cite{Greensite03,Greensite04,GreensiteThorn}. 
  On the other hand, in the Landau gauge, the instantaneous potential has no linear part, and 
  the ghost behavior in the deep-infrared region would be more important for the confinement
  \cite{Alkofer01,FuruiNakajima}.

  \subsection{``Finite-time Q$\bar{\bf Q}$ potential''}

  In the previous subsection, we investigated 
  the instantaneous potential $V_\lambda(R)$, which is defined
  by the Polyakov-line correlator with a minimum length on the lattice.
  We investigate the ``finite-time potential'' $V_\lambda(R,T)$ defined by 
  Eq.(\ref{eqFTPot}) in Sec.IV,
  and its temporal-length dependence.  
  Here, $V_\lambda(R,T)$ is expressed by $T$-length terminated Polyakov-line 
  $L(\vec{x},T)$ in Eq.(\ref{eqterminatedPL}),
  and a generalization of the instantaneous potential $V_\lambda(R)$.

\begin{figure}
    \centering
    \includegraphics[width=8cm,clip]{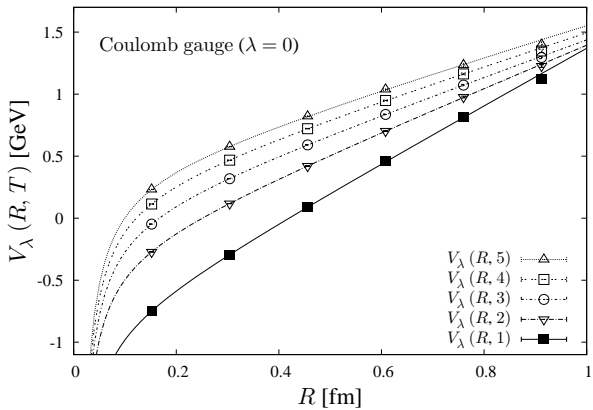}
    \caption{\label{figExtendedPot}
      ``Finite-time potential'' $V_\lambda(R,T)$
      in the Coulomb gauge ($\lambda = 0$) for $T$=1,2,3,4,5.
      Here, for the comparison, an irrelevant constant is shifted for each $T$.
      The curves denote the fit-results using the Coulomb plus linear form.
      The slope of $V_\lambda(R,T)$ is clearly changed according to $T$.
    }
  \end{figure}

  First, we consider the Coulomb gauge \cite{Greensite03,Iritani10}.
  Figure \ref{figExtendedPot} shows 
  the lattice QCD result for $V_\lambda(R,T)$ in the Coulomb gauge. 
  Similar to the instantaneous potential,
  $V_\lambda(R,T)$ is well reproduced by the Coulomb plus linear form.
  However, the parameter values are changed according to $T$-length.
  In particular, the slope of the potential becomes smaller as $T$ becomes larger.

  For general $\lambda$, 
  finite-time potential $V_\lambda(R,T)$ is found to be reproduced 
  by the Coulomb plus linear form as 
  \begin{equation}
    V_\lambda(R,T) 
    = - \frac{A_\lambda(T)}{R} + \sigma_\lambda(T) R + C_\lambda(T),
  \end{equation}
  at least for $R \lesssim 0.8$fm, similarly for the instantaneous potential. 

\begin{figure}
    \centering
    \includegraphics[width=8cm,clip]{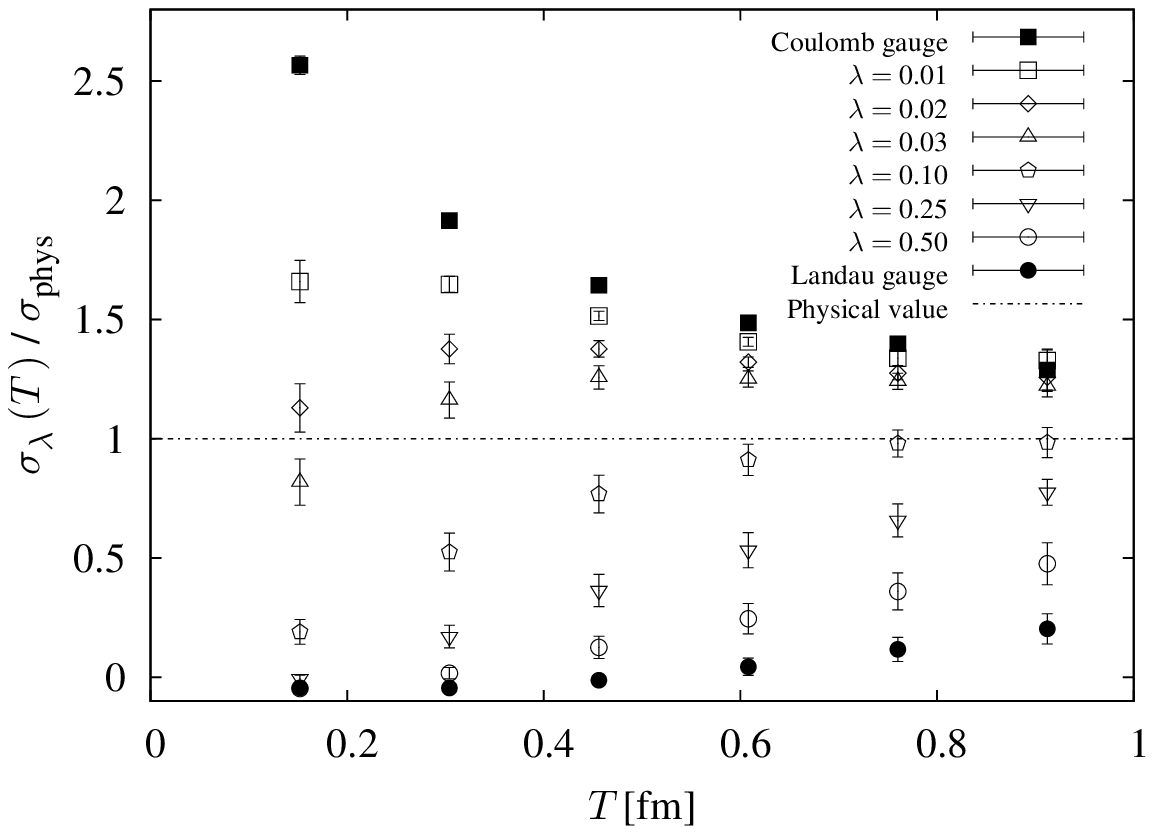}
    \caption{\label{figStringTensionLengthDep}
     $T$-length dependence of ``Finite-time string tension'' $\sigma_\lambda(T)$, 
     the infrared slope of finite-time potential $V_\lambda(R,T)$, 
     in generalized Landau gauge for several typical $\lambda$-values.
     Near the Coulomb gauge, e.g., for $\lambda \lesssim 0.03$,
     $\sigma_\lambda(T)$ goes to the same value for large $T \sim$ 1fm.
    For $\lambda \gtrsim 0.1$, $\sigma_\lambda(T)$ is an increasing function of $T$.
    In fact, even though the instantaneous potential has no linear part, 
    the linear part of $V_\lambda(R,T)$ appears and gradually grows, 
    as the Polyakov-line grows.
    }
  \end{figure}

  Here, we investigate ``finite-time string tension'' $\sigma_\lambda(T)$, 
  the slope of $V_\lambda(R,T)$.
  Figure \ref{figStringTensionLengthDep} shows 
  $\sigma_\lambda(T)$ in generalized Landau gauge for typical values of $\lambda$.
  In Table \ref{tabTlengthStringTension}, 
  we summarize the best-fit parameters of 
  $\sigma_\lambda(T)$ at $T = 1, 2, \dots, 6$,
  and the ratio of 
  $\sigma_\lambda(1)/\sigma_\lambda(6)$,    
  $\sigma_\lambda(1)/\sigma_{\rm phys}$, and
  $\sigma_\lambda(6)/\sigma_{\rm phys}$, respectively.

  \begin{table*}
    \begin{center}
    \caption{\label{tabTlengthStringTension}
      Finite-time string tension $\sigma_\lambda(T)$ in generalized Landau gauge
      for $T = 1, 2, \dots, 6$, together with the ratio,  
      $\sigma_\lambda(1)/\sigma_\lambda(6)$,    
      $\sigma_\lambda(1)/\sigma_{\rm phys}$, and
      $\sigma_\lambda(6)/\sigma_{\rm phys}$.
      The fit-range is the same as that listed in Table \ref{tabFitResult}.
    }
    \begin{tabular}{cccccccccc}
      \hline
      \hline
      & \multicolumn{6}{c}{$\sigma_\lambda(T)$ [GeV/fm]} \\
      $\lambda$ & $T=1$ & $T=2$ & $T=3$ & $T=4$ & $T=5$ & $T=6$ &
      $\sigma_\lambda(1)/\sigma_\lambda(6)$ &
      $\sigma_\lambda(1)/\sigma_{\rm phys}$ &
      $\sigma_\lambda(6)/\sigma_{\rm phys}$ \\
      \hline
      0.00  &  2.283(35) & 1.704(11) & 1.463(8) & 1.322(16) &  1.244(24) & 1.147(75) & 1.99(13) &   2.57(4) &  1.29(8) \\
      0.01  &  1.476(78) & 1.466(30) & 1.348(17) & 1.252(16) &  1.191(27) & 1.181(43) & 1.25(8) &   1.66(9) &  1.33(5) \\
      0.02  &  1.005(90) & 1.225(55) & 1.225(31) & 1.176(20) &  1.135(25) & 1.119(53) & 0.90(9) &   1.13(10)&  1.26(6) \\
      0.03  &  0.728(86) & 1.034(67) & 1.119(43) & 1.113(30) &  1.104(30) & 1.086(40) & 0.67(8) &   0.82(10)&  1.22(5) \\
      0.04  &  0.557(79) & 0.896(72) & 1.030(51) & 1.057(35) &  1.053(30) & 1.019(53) & 0.55(8) &   0.63(9) &  1.15(6) \\
      0.05  &  0.441(73) & 0.785(76) & 0.947(59) & 1.002(43) &  1.021(35) & 1.020(44) & 0.43(7) &   0.50(8) &  1.15(5) \\
      0.10  &  0.169(46) & 0.467(71) & 0.684(70) & 0.811(58) &  0.872(50) & 0.875(56) & 0.19(5) &   0.19(5) &  0.98(6) \\
      0.25  &  -0.005(13) & 0.152(42) & 0.324(60) & 0.474(65) &  0.586(62) & 0.690(48) & -0.01(2) & -0.01(1) &  0.78(5) \\
      0.50  &  -0.042(1) & 0.015(21) & 0.111(41) & 0.218(57) &  0.320(69) & 0.423(79) & -0.10(2) & -0.05(0)  &  0.48(9) \\
      0.75  &  -0.043(3) & -0.025(11) & 0.028(27) & 0.100(44) &  0.181(58) & 0.275(69) & -0.16(4) & -0.05(0)  &  0.31(8) \\
      1.00  &  -0.040(3) & -0.041(5) & -0.012(18) & 0.039(32) &  0.104(45) & 0.180(56) & -0.22(7) & -0.04(0) &  0.20(6) \\
      \hline
      \hline
    \end{tabular}
    \end{center}
  \end{table*}

  On the $T$-dependence of the finite-time string tension $\sigma_\lambda(T)$,
  there are three groups of the $\lambda$-parameter region:
  (i) Coulomb-like region ($0 \le \lambda \ll \lambda_C$),
  (ii) Landau-like region ($\lambda_C \ll \lambda \le 1$), and  
  (iii) $\lambda_C$-like region ($\lambda \sim \lambda_C$).

  (i) The first category is the Coulomb-like region of $0 \le \lambda \ll \lambda_C$, 
  i.e., $0 \le \lambda \lesssim 0.01$. 
  In this region, the instantaneous string tension $\sigma_\lambda \equiv \sigma_\lambda(T=1)$
  is larger than the physical string tension $\sigma_{\rm phys}$, and 
  the instantaneous potential $V_\lambda(R) \equiv V_\lambda(R,T=1)$ gives an overconfining potential. 
  The ground-state of the quark-antiquark system is considered as the gluon-chain state.
  As the temporal length $T$ of the Polyakov-line increases, finite-time string tension 
  $\sigma_\lambda(T)$ decreases and approaches to the physical string tension $\sigma_{\rm phys}$.
  This decreasing behavior is interpreted that 
  the component of the ground-state, i.e., the gluon-chain state, 
  becomes dominant as $T$-length becomes large.

  (ii) The second category is the Landau-like region of $\lambda_C \ll \lambda \le 1$, 
  i.e. $0.1 \lesssim \lambda \le 1$. 
  In this region, the instantaneous string tension is almost zero, 
  i.e., $\sigma_\lambda \simeq 0$,
  and finite-time string tension $\sigma_\lambda(T)$ is an increasing function of $T$.
  Although its asymptotic value is unclear for $T \sim 0.8$ fm,
  $\sigma_\lambda(T)$ seems to approach to the physical string tension $\sigma_{\rm phys}$,
  which will be discussed in Sec.VI.

\begin{figure}
    \centering
    \includegraphics[width=8cm,clip]{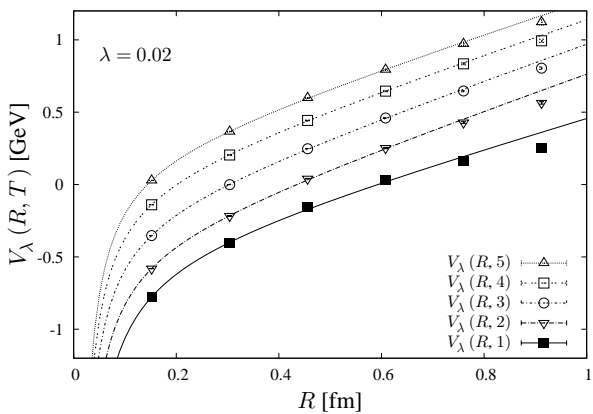}
    \caption{\label{figExtendedPotLambdaC}
      Finite-time potential $V_\lambda(R,T)$ at $\lambda = 0.02(\simeq \lambda_C)$.
      For the comparison, an irrelevant constant is shifted for each $T$.
      The slope of $V_\lambda(R,T)$ is almost the same for $T=1,2, \cdots 5$.
    }
  \end{figure}

  (iii) The third category is the $\lambda_C$-like region of $\lambda \sim \lambda_C$, 
  i.e., $0.01 \lesssim \lambda \lesssim 0.1$.
  In this region, the instantaneous string tension $\sigma_\lambda$ 
  is approximately equal to the physical string tension, i.e., 
  $\sigma_\lambda \simeq \sigma_{\rm phys} (\simeq$ 0.89GeV/fm), 
  and the instantaneous potential $V_\lambda(R)$ approximately reproduces 
  the physical static potential $V_{\rm phys}(R)$ for $R \lesssim 0.8$fm.
  As the temporal length $T$ increases, 
  finite-time string tension $\sigma_\lambda(T)$  
  is slightly changed and takes a little larger value ($\simeq$ 1.1GeV/fm) 
  around $T \simeq$ 0.8fm.
  In particular, near $\lambda_C \simeq 0.02$, 
  $\sigma_\lambda(T)$ shows only a weak $T$-dependence,   
  while $\sigma_\lambda(T)$ largely changes as $T$ in the Coulomb gauge. 
  As a whole, finite-time potential $V_\lambda(R,T)$ has small $T$-dependence, 
  as shown in Fig.~\ref{figExtendedPotLambdaC}.

  When the length $T$ of the Polyakov-line increases,
  $\lambda$-dependence of $\sigma_\lambda(T)$ is weakened,
  and $\sigma_\lambda(T)$ seems to converge to  
  the physical string tension $\sigma_{\rm phys}$ for enough large $T$,
  as indicated in Fig.\ref{figStringTensionLengthDep}.

  There are two ingredients on the above gauge-dependence ($\lambda$-dependence):
  one is the large excess of the Coulomb energy in the Coulomb gauge,   
  the other is the non-locality from the Faddeev-Popov determinant.
  From the fixing condition of generalized Landau gauge 
  in Eq.(\ref{eqlambdafix}), one finds that 
  the $\lambda$-parameter controls the non-locality in the temporal direction.
  In the Landau gauge, the non-locality appears 
  equally in spatial and temporal directions,
  while temporal non-locality disappears in the Coulomb gauge.
  If the Coulomb-energy excess can be neglected, e.g., for large $\lambda$, 
  $V_\lambda(R,T)$ is expected to reproduce the static potential $V_{\rm phys}(R)$,
  when $T$-length and $R$ are large enough to neglect the non-locality scale. 
  Owing to the $\lambda$-dependence of the non-locality, 
  such a $T$-length exceeding the non-locality is to be larger 
  for larger $\lambda$ in the Landau-like region.

  \subsection{``Instantaneous 3Q inter-quark potential'' in generalized Landau gauge}

\begin{figure}
    \centering
    \includegraphics[width=8cm,clip] {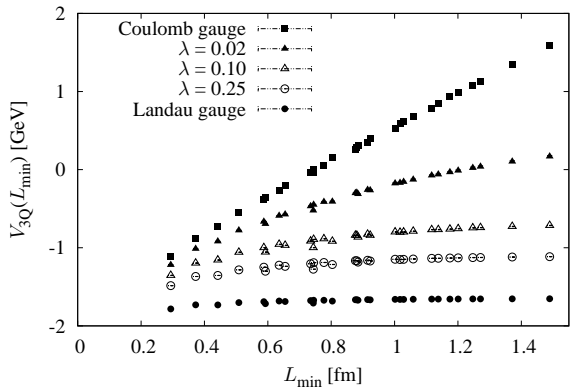}
    \caption{\label{figInstPot3Q}
      ``Instantaneous potential''
      $V_\lambda^{\rm 3Q}$ plotted against $L_{\rm min}$ 
      in generalized Landau gauge
      for typical values of $\lambda$.
      The symbols denote lattice QCD results.
      For $\lambda = 1 \sim 0.1$, 
      the potential has almost no linear part.
      For $\lambda \lesssim 0.1$,
      the linear potential grows rapidly, and 
      $\sigma_\lambda^{\rm 3Q} \simeq 2.5\sigma_{\rm phys}$ at $\lambda = 0$.
    }
  \end{figure}

  We investigate also the instantaneous 3Q potential 
  $V_\lambda^{\rm 3Q}$ in generalized Landau gauge.
  Figure \ref{figInstPot3Q} shows lattice QCD results
  of $V_\lambda^{\rm 3Q}$ for typical values of $\lambda$.
  In Fig.\ref{figInstPot3Q} $V_\lambda^{\rm 3Q}$ is plotted against the minimal
  value of the total flux-tube length, $L_{\rm min}$.
  Using the length of the sides of 3Q triangle, $a,b$ and $c$, $L_{\rm min}$ is expressed as
  \begin{widetext}
    \begin{equation}
      L_{\rm min} = \left[ \frac{1}{2} (a^2+b^2+c^2) +
      \frac{\sqrt{3}}{2} \sqrt{(a+b+c)(a-b+c)(a+b-c)(-a+b+c)}\right]^{1/2},
    \end{equation}
  \end{widetext}
  when all angles of the 3Q triangle do not exceed $2\pi/3$, 
  and
  \begin{equation}
    L_{\rm min} = a + b + c - \mathrm{max}(a,b,c),
  \end{equation}
  when an angle of the 3Q triangle exceeds $2\pi/3$ \cite{STI,TS0304}.
  For the ground-state physical 3Q potential, 
  recent lattice QCD studies have shown that 
  the confinement part is proportional to $L_{\rm min}$, 
  which is called as Y-Ansatz \cite{STI,TS0304,Rothe,Ichie}.

  While there is no linear part at all in the Landau gauge ($\lambda = 1$), 
  the instantaneous potential shows linear behavior proportional to $L_{\rm min}$ 
  in the Coulomb gauge ($\lambda = 0$), 
  which indicates the Y-Ansatz even for the instantaneous potential \cite{BS04}.

  By varying the $\lambda$-parameter from 1 to 0 in the generalized Landau gauge, 
  we find also that the instantaneous 3Q potential $V_\lambda^{\rm 3Q}$ changes continuously, and  
  the infrared slope of the potential $V_\lambda^{\rm 3Q}$ 
  at $L_{\rm min} \simeq 1.5 {\rm fm}$ grows monotonically,
  from the Landau gauge to the Coulomb gauge, as shown in Fig.\ref{figInstPot3Q}.
  Note here that 
  the growing of the infrared slope of $V_\lambda^{\rm 3Q}$ is quite rapid
  for $\lambda \lesssim 0.1$ near the Coulomb gauge,
  while the infrared slope is rather small and almost unchanged for $\lambda = 0.1 \sim 1$.
  This gauge dependence of the potential is similar to Q$\bar{\rm Q}$ potential
  $V_\lambda(R)$.

   To analyze the instantaneous 3Q potential $V_\lambda^{\rm 3Q}$ quantitatively,
   we fit the lattice QCD results using 
   the Coulomb plus linear Ansatz as
   \begin{equation}
     V_\lambda^{\rm 3Q}(\vec{x}_1, \vec{x}_2, \vec{x}_3) 
       = - A_\lambda^{\rm 3Q} \sum_{i<j}\frac{1}{|\vec{x}_i-\vec{x}_j|} 
       + \sigma_\lambda^{\rm 3Q} L_{\rm min} + C_\lambda^{\rm 3Q},
     \label{eqCoulplusLin3Q}
   \end{equation} 
   where $A_\lambda^{\rm 3Q}$ is the Coulomb coefficient 
   and $C_\lambda^{\rm 3Q}$ a constant.
   Here, $\sigma_\lambda^{\rm 3Q}$ is the infrared slope of the potential, 
   which we call as ``instantaneous string tension''.

   We summarize the best-fit parameters and the fit-range in Table \ref{tabFitResult3Q}.
   While the Coulomb coefficient $A^{\rm 3Q}_\lambda$ has 
   a relatively weak $\lambda$-dependence, 
   the instantaneous string tension 
   $\sigma^{\rm 3Q}_\lambda$ shows a strong $\lambda$-dependence near 
   the Coulomb gauge, i.e., $\lambda \lesssim 0.1$.

   Similar to the Q$\bar{\rm Q}$ case, 
   in the deep-IR limit, $L_{\rm min} \rightarrow \infty$, $V_\lambda^{\rm 3Q}$
   goes to a saturated value, except for $\lambda = 0$.
   This behavior is also due to $\langle U_4\rangle \ne 0$ for the temporal 
   link-variable $U_4$ in the generalized Landau gauge for $\lambda \ne 0$,
   as will be discussed in Sec.VI.

  \begin{table}
    \begin{center}
    \caption{\label{tabFitResult3Q}
      The best-fit parameters on the instantaneous potential using
      $V_\lambda^{\rm 3Q}(\vec{x}_1, \vec{x}_2, \vec{x}_3) 
      = - A_\lambda^{\rm 3Q} \sum_{i<j}\frac{1}{|\vec{x}_i-\vec{x}_j|} 
      + \sigma_\lambda^{\rm 3Q} L_{\rm min} + C_\lambda^{\rm 3Q}$,
      and the ratio of the slope $\sigma_\lambda^{\rm 3Q}$ to the physical string tension 
      $\sigma_{\rm phys}$.
      Fit-range is $L_{\rm min} = 0.3 \sim 1.5$fm.
      The standard parameters of the physical interquark potential 
      are $A^{\rm 3Q}_{\rm phys} \simeq A^{\rm Q\bar{\rm Q}}_{\rm phys}/2$, 
      ($A^{\rm Q\bar{\rm Q}}_{\rm phys} \simeq 0.27$) 
      and $\sigma_{\rm phys} \simeq 0.89$GeV/fm \cite{STI}.
      The string tension $\sigma^{\rm 3Q}_\lambda$ is rather small for $\lambda = 0.1 \sim 1$.
    }
    \begin{tabular}{ccccc}
    \hline
    \hline 
    $\lambda$ & $A_\lambda^{\rm 3Q}$ & $\sigma_\lambda^{\rm 3Q}$ [GeV/fm] 
      & $C_\lambda^{\rm 3Q}$ [GeV] & $\sigma_\lambda^{\rm 3Q}/\sigma_{\rm phys}$ \\
    \hline
    0.00  & 0.028(13) &  2.196(39)  & -1.617(50)  & 2.47(4)  \\ 
    0.01  & 0.165(15) &  1.192(41)  & -0.954(55)  & 1.34(5)  \\
    0.02  & 0.211(14) &  0.739(35)  & -0.709(48)  & 0.83(4)  \\ 
    0.03  & 0.223(12) &  0.503(30)  & -0.626(42)  & 0.57(3)  \\ 
    0.04  & 0.227(10) &  0.363(25)  & -0.593(36)  & 0.41(3)  \\ 
    0.05  & 0.225(9)  &  0.275(22)  & -0.590(32)  & 0.31(3)  \\ 
    0.10  & 0.199(6)  &  0.084(13)  & -0.685(19)  & 0.09(1)  \\ 
    0.25  & 0.137(4)  & -0.018(8)   & -0.991(13)  &-0.02(1)  \\ 
    0.50  & 0.090(3)  & -0.031(7)   & -1.288(11)  &-0.04(1)  \\ 
    0.75  & 0.067(3)  & -0.028(6)   & -1.463(10)  &-0.03(1)  \\ 
    1.00  & 0.054(3)  & -0.026(5)   & -1.579(8)   &-0.03(1)  \\ 
    \hline 
    \hline
    \end{tabular}
    \end{center}
  \end{table}

\begin{figure}
    \centering
    \includegraphics[width=8cm,clip] {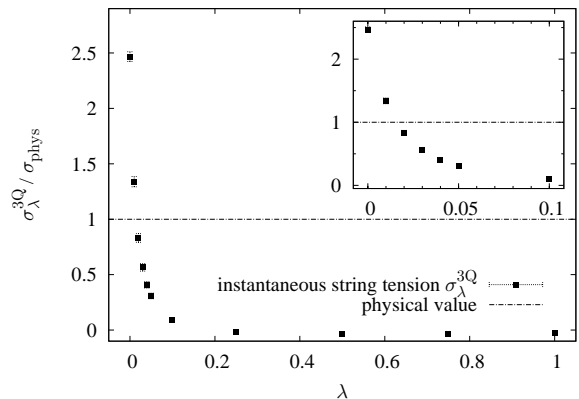}
    \caption{ \label{figStringTension3Q}
      Instantaneous string tension $\sigma_\lambda^{\rm 3Q}$, the slope of 
      the linear part in the instantaneous potential $V_\lambda^{\rm 3Q}$
      in the generalized Landau gauge.
      The right upper figure is a close-up near the Coulomb gauge ($\lambda$=0).
      For $0 \le \lambda \lesssim 0.1$,  
      $\sigma^{\rm 3Q}_\lambda$ changes rapidly from $2.5\sigma_{\rm phys}$ to 0,
      while $\sigma^{\rm 3Q}_\lambda$ is rather small for $0.1 \lesssim \lambda \le 1$.
    }
  \end{figure}

  Figure \ref{figStringTension3Q} shows
  the $\lambda$-dependence of instantaneous string tension $\sigma_\lambda^{\rm 3Q}$.
  Similarly to the Q$\bar{\rm Q}$ case,
  $\sigma_\lambda^{\rm 3Q}$ is almost zero
  in Landau-like region ($0.1 \lesssim \lambda \leq 1$),
  and $\sigma_\lambda^{\rm 3Q}$ grows rapidly in Coulomb-like region 
  ($0 \leq \lambda \lesssim 0.1$).
  Finally, in the Coulomb gauge ($\lambda$=0), one finds 
  $\sigma_\lambda^{\rm 3Q} \simeq 2.5 \sigma_{\rm phys}$, 
  with $\sigma_{\rm phys} \simeq 0.89$GeV/fm.

  Similarly to the instantaneous string tension for the Q$\bar{\rm Q}$ system,
  $\sigma_\lambda^{\rm 3Q}$ continuously changes from 0 to $(2 \sim 3)\sigma_{\rm phys}$
  by varying from the Landau gauge to the Coulomb gauge. (See also Fig.\ref{figStringTension}.)
  Therefore, there exists some specific $\lambda$-parameter of $\lambda_C \in [0,1]$
  where the slope of the instantaneous potential $V_\lambda^{\rm 3Q}$
  coincides with the physical string tension $\sigma_{\rm phys}$.
  For the 3Q system, we find $\lambda_C \simeq$ 0.02, 
  which is almost the same value for the Q$\bar{\rm Q}$ case. 
  Then, as a whole, a universal behavior is found for the instantaneous string tension 
  between Q$\bar{\rm Q}$ systems and 3Q systems.

  \subsection{``Finite-time 3Q potential''}

\begin{figure}
    \centering
    \includegraphics[width=8cm,clip] {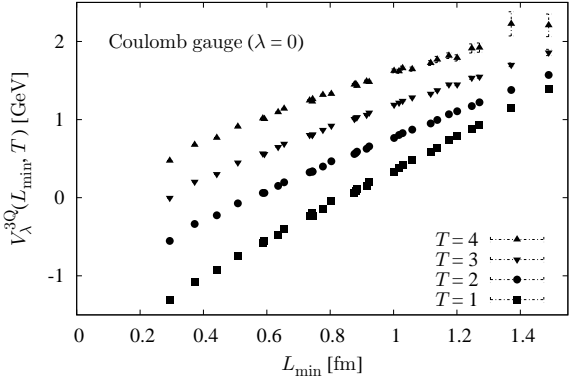}
      \caption{\label{figExtendedPot3Q}
        ``Finite-time 3Q potential'' $V_\lambda^{\rm 3Q}(\vec{x}_1, \vec{x}_2,\vec{x}_3,T)$
        plotted against $L_{\rm min}$
        in the Coulomb gauge ($\lambda = 0$) for $T$=1,2,3,4.
        Here, for the comparison, an irrelevant constant is shifted for each $T$.
    }
  \end{figure}

  Next, we consider the finite-time 3Q potential.
  Figure~\ref{figExtendedPot3Q} shows the lattice QCD result
  for $V_\lambda^{\rm 3Q}$ in the Coulomb gauge,
  which is well reproduced by the plus linear form.

  For general $\lambda$, 
  finite-time potential $V_\lambda^{\rm 3Q}(T)$ is found to be reproduced 
  by the Coulomb plus linear form as 
  \begin{eqnarray}
    V_\lambda^{\rm 3Q}(\vec{x}_1,\vec{x}_2,\vec{x}_3,T)  
     &=& -{A_\lambda^{\rm 3Q}(T)}\sum_{i<j}\frac1{|\vec{x}_i-\vec{x}_j|} 
     \nonumber \\
     &&\ + \sigma_\lambda^{\rm 3Q}(T) L_{\rm min} + C_\lambda^{\rm 3Q}(T),
  \end{eqnarray}
  at least for $L_{\rm min} \lesssim 1.5$fm.

\begin{figure}
      \centering
      \includegraphics[width=8cm,clip] {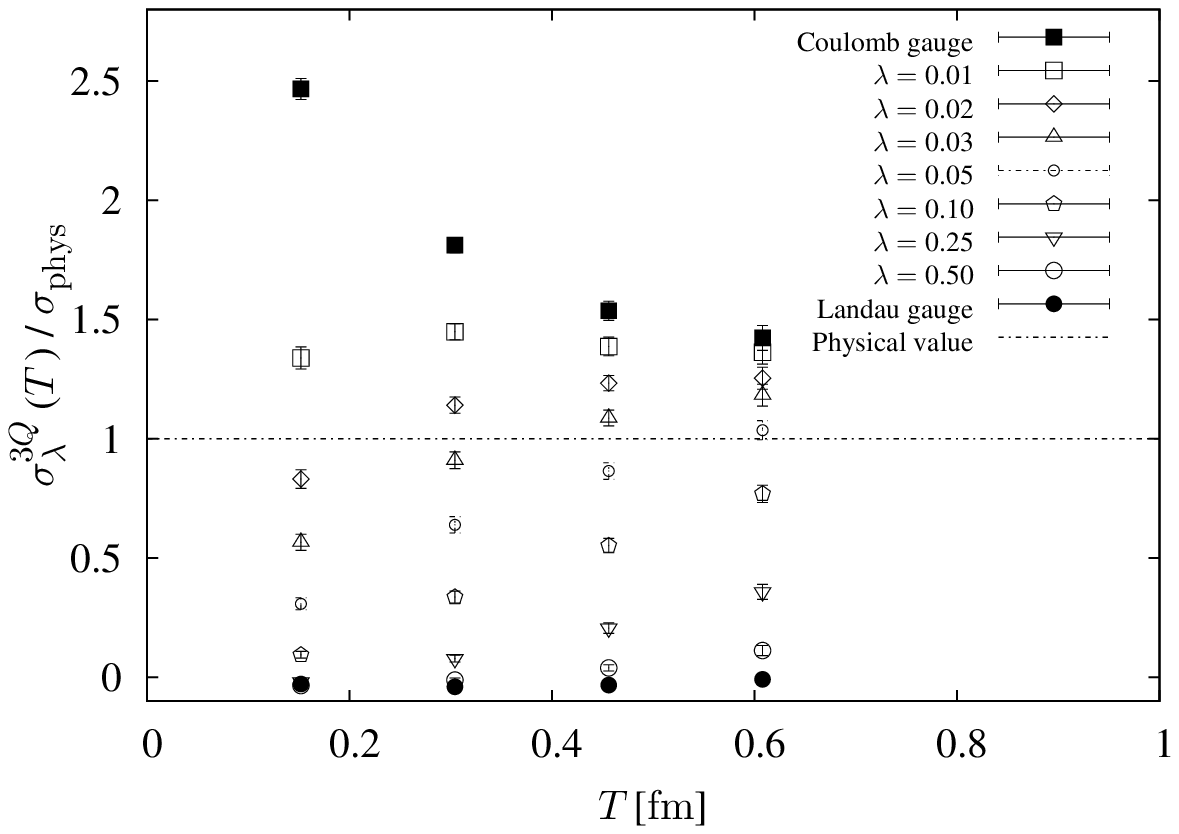}
      \caption{\label{fig3StringTensionLengthDep}
        $T$-length dependence of ``Finite-time string tension'' $\sigma_\lambda^{\rm 3Q}(T)$, 
        the infrared slope of finite-time 3Q potential $V_\lambda^{\rm 3Q}(T)$,
        in generalized Landau gauge for several typical $\lambda$-values.
        Near the Coulomb gauge, e.g., for $\lambda \lesssim 0.03$,
        $\sigma_\lambda^{\rm 3Q}(T)$ goes to the same value for large $T \sim$ 0.6fm.
        For $\lambda \gtrsim 0.1$, $\sigma_\lambda^{\rm 3Q}(T)$ is 
        an increasing function of $T$.
        In fact, even though the instantaneous potential has no linear part, 
        the linear part of $V_\lambda^{\rm 3Q}(T)$ appears and gradually grows, 
        as the Polyakov-line grows.
      }
    \end{figure}

  Here, we investigate ``finite-time string tension'' $\sigma^{\rm 3Q}_\lambda(T)$, 
  the slope of $V_\lambda^{\rm 3Q}(T)$.
  Figure \ref{fig3StringTensionLengthDep} shows 
  $\sigma^{\rm 3Q}_\lambda(T)$ in generalized Landau gauge
  for typical values of $\lambda$.
  In Table \ref{tab3TlengthStringTension}, 
  we summarize the best-fit parameters of 
  $\sigma^{\rm 3Q}_\lambda(T)$ at $T = 1, 2, 3, 4$.
  For $T \gtrsim 5$, the meaningful fit result was not obtained due to large errors.

\begin{figure}
      \centering
      \includegraphics[width=8cm,clip] {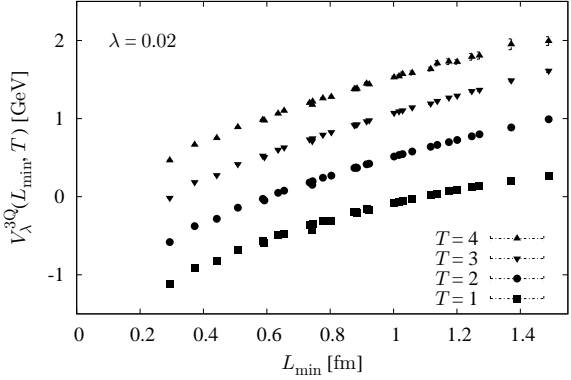}
      \caption{\label{figExtendedPotLambdaC_3Q}
        Finite-time potential $V_\lambda^{3Q}(\vec{x}_1,\vec{x}_2,\vec{x}_3,T)$ 
        plotted against $L_{\rm min}$ at $\lambda = 0.02(\simeq \lambda_C)$.
        For the comparison, an irrelevant constant is shifted for each $T$.
        The slope of $V_\lambda^{3Q}(L_{\rm min},T)$ 
        is approximately the same for $T=1,2,3,4$.
      }
    \end{figure}

  The behavior of $\sigma^{\rm 3Q}_\lambda(T)$ 
  is similar to the Q$\bar{\rm Q}$ case.
  (i) In the Coulomb-like region ($\lambda \simeq 0$), 
  $\sigma_\lambda^{\rm 3Q}(T)$ is a decreasing function, and approaches to 
  the physical string tension $\sigma_{\rm phys}$ as $T$ increases.
  (ii) In the Landau-like region ($0.1 \lesssim \lambda \le 1$), 
  $\sigma_\lambda^{\rm 3Q}(T)$ is an increasing function of $T$: 
  starting from zero at $T=1$, 
  the linear part of $V_\lambda^{\rm 3Q}(T)$ appears and grows, 
  as $T$ increases.
  (iii) In the $\lambda_C$-like region ($\lambda \sim \lambda_C\simeq 0.02)$, 
  $T$-dependence is relatively weak, as shown in Fig.\ref{figExtendedPotLambdaC_3Q}.

  At the quantitative level,  
  the 3Q finite-time string tension $\sigma^{\rm 3Q}_\lambda(T)$ 
  shows the similar behavior as 
  the Q$\bar{\rm Q}$ finite-time string tension $\sigma_\lambda(T)$.
  (See Table.\ref{tabTlengthStringTension}
  and Fig.\ref{figStringTensionLengthDep}.)
  In fact, the finite-time string tension shows a universal behavior on the $T$-dependence  
  for both Q$\bar{\rm Q}$ systems and 3Q systems.

    \begin{table} 
      \begin{center}
      \caption{\label{tab3TlengthStringTension}
        Finite-time string tension $\sigma_\lambda^{\rm 3Q}(T)$ in generalized Landau gauge
        for $T = 1, 2, 3, 4$.
        The fit-range is the same as that listed in Table \ref{tabFitResult3Q}.
      }
      \begin{tabular}{ccccc}
        \hline
        \hline
          & \multicolumn{4}{c}{$\sigma_\lambda^{\rm 3Q}(T)$ [GeV/fm]} \\
          $\lambda$ & $T=1$ & $T=2$ & $T=3$ & $T=4$  \\
        \hline
        0.00 &  2.196(39) & 1.612(28) & 1.368(35) & 1.266(46)  \\
        0.01 &  1.192(41) & 1.290(30) & 1.234(35) & 1.213(45)  \\
        0.02 &  0.739(35) & 1.016(30) & 1.097(28) & 1.116(41)  \\
        0.03 &  0.503(30) & 0.810(31) & 0.967(29) & 1.053(41)  \\
        0.04 &  0.363(25) & 0.670(31) & 0.855(31) & 0.989(39)  \\
        0.05 &  0.275(22) & 0.569(30) & 0.770(31) & 0.922(36)  \\
        0.10 &  0.084(13) & 0.299(22) & 0.491(27) & 0.685(32)  \\
        0.25 &  -0.018(8) & 0.070(12) & 0.183(19) & 0.319(27)  \\
        0.50 &  -0.031(7) & -0.011(8) & 0.035(12) & 0.100(19)  \\
        0.75 &  -0.028(6) & -0.029(7) & -0.008(9) & 0.029(14)  \\
        1.00 &  -0.026(5) & -0.036(7) & -0.030(8) & -0.008(10) \\
        \hline
        \hline
      \end{tabular}
      \end{center}
    \end{table}

\section{Terminated Polyakov-line correlator and potentials}

  In the previous section, for both Q$\bar{\rm Q}$ and 3Q systems, 
  we investigated the instantaneous potential ($V_\lambda(R)$, $V_\lambda^{\rm 3Q}$) 
  and the finite-time potential ($V_\lambda(R,T)$, $V_\lambda^{\rm 3Q}(T)$), 
  which are derived from the correlation of terminated Polyakov-line $L(\vec{x},T)$.
  In this section, we investigate properties of the Polyakov-line correlator, 
  and clarify its relation to the instantaneous/finite-time potential
  for both Q$\bar{\rm Q}$ and 3Q systems.

  \subsection{Asymptotic behavior of link-variable correlator and instantaneous potential}
  First, we investigate the spatial correlator $G_\lambda(R)$ of 
  the temporal link-variable $U_4$,
  and the relation to the instantaneous potential 
  $V_\lambda(R) \equiv -\frac{1}{a}{\rm ln} G_\lambda(R)$.
  For the large spatial separation of $R \equiv |\vec{x}-\vec{y}| \rightarrow \infty$,  
  $G_\lambda(R)$ behaves asymptotically as
  \begin{eqnarray}
    G_\lambda(R) &\equiv& 
    \langle \mathrm{Tr} \ [U_4^\dagger(\vec{x},t) U_4(\vec{y},t)] \rangle \nonumber \\
    &\rightarrow& 
    \langle (U_4)_{ij}^* \rangle \langle (U_4)_{ij} \rangle 
    = \frac{1}{3} \langle \mathrm{Tr} \ U_4 \rangle^2,
  \end{eqnarray}
  where
  $\langle (U_4)_{ij}\rangle=\frac{1}{3}\langle \mathrm{Tr} \ U_4 \rangle \delta_{ij} 
  \in \mathbf{R}$ from the global color symmetry.
  Here, $\frac{1}{3}\langle \mathrm{Tr} \ U_4 \rangle^2$ is found to  
  give the lower bound of $G_\lambda(R)$.
  If $\langle \mathrm{Tr} \ U_4\rangle$ takes some finite value,
  $V_\lambda(R)$ inevitably saturates for large $R$. 
  Then, $\langle \mathrm{Tr} \ U_4\rangle=0$ is a necessary condition for 
  the deep-infrared confinement feature of 
  $V_\lambda(R=\infty) = \infty$.      

  In the Coulomb gauge,
  $\langle \mathrm{Tr} \ U_4 \rangle$ is zero due to the remnant symmetry, 
  as was shown in Sec.III B.
  Therefore, as $R \rightarrow \infty$, the correlator $G_\lambda(R)$ converges to zero, 
  and $V_\lambda(R)\equiv -\frac{1}{a}{\rm ln} G_\lambda(R) \rightarrow +\infty$, 
  which corresponds to the deep-infrared confinement.

  For the general case of $\lambda \neq 0$, however, 
  $\langle \mathrm{Tr} \ U_4 \rangle$ has a non-zero value, 
  and $G_\lambda(R)$ approaches to a non-zero finite constant.
  The finiteness of $\langle \mathrm{Tr} \ U_4 \rangle$
  gives a saturation of $V_\lambda(R)$,  
  which leads to the absence of its linear part in the deep-IR region. 

\begin{figure}
    \centering
    \includegraphics[width=8cm,clip] {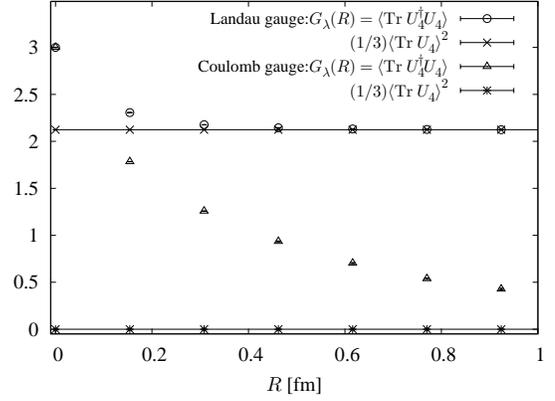}
    \caption{ \label{figU4Corr}
      The spatial correlator $G_\lambda(R) \equiv 
        \langle \mathrm{Tr} \ U_4^\dagger(\vec{x},a) U_4(\vec{y},a)\rangle$
      $(R = |\vec{x}-\vec{y}|)$  
      in the Landau gauge (open circles) and the Coulomb gauge (open triangles), 
      together with its asymptotic value of 
      $\frac{1}{3}\langle\mathrm{Tr} \ U_4 \rangle^2$ (solid lines and cross symbols).        
      In the Landau gauge, $\langle \mathrm{Tr} \ U_4\rangle \ne 0$, 
      and $G_\lambda(R)$ rapidly converges to a constant for $R \gtrsim$ 0.4fm,
      which leads to a rapid saturation of the instantaneous potential 
      $V_\lambda(R)\equiv -\frac{1}{a}{\rm ln} G_\lambda(R)$.
      In the Coulomb gauge, 
      $\langle \mathrm{Tr} \ U_4 \rangle = 0$, and
      $G_\lambda(R)$ decreases monotonically to zero as $R$, 
      which leads to $V_\lambda(R)=+\infty$ for $R=\infty$.
    }
  \end{figure}

  Figure \ref{figU4Corr} shows $G_\lambda(R)$ and its asymptotic value 
  $\frac{1}{3} \langle \mathrm{Tr} \ U_4\rangle^2$ 
  in the Landau and the Coulomb gauges.
  In the Landau gauge, $\langle \mathrm{Tr} \ U_4 \rangle$ has a large expectation value,
  according to the maximization of $\sum_x \mathrm{Re} \ \mathrm{Tr} \ U_\mu(x)$,
  and $G_\lambda(R)$ rapidly converges to a finite constant for $R \gtrsim$ 0.4fm, 
  which leads to a rapid saturation of the instantaneous potential $V_\lambda(R)$.
  In the Coulomb gauge, we find 
  $\langle \mathrm{Tr} \ U_4 \rangle = 0$,  
  and $G_\lambda(R)$ decreases monotonically to zero as $R$,
  which leads to $V_\lambda(R)=+\infty$ for $R=\infty$.

\begin{figure}
    \centering
    \includegraphics[width=8cm,clip]{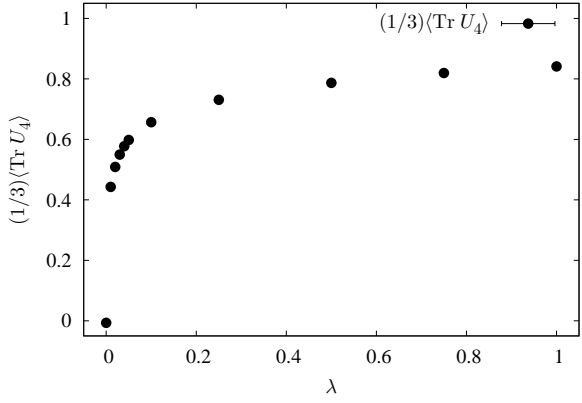}
    \caption{ \label{figTrU4}
       The expectation value of $\frac{1}{3}\langle \mathrm{Tr} \ U_4 \rangle$ 
       in generalized Landau gauge.
       For $\lambda \ne 0$, 
       $\langle \mathrm{Tr} \ U_4 \rangle$ takes a non-zero value,
       and it approaches to zero continuously as $\lambda \rightarrow 0$.
       The value of $\langle \mathrm{Tr} \ U_4 \rangle$ relates to  
       the infrared behavior of the correlator $G_\lambda(R)$ 
       and the instantaneous potential $V_\lambda(R)$.
    }
  \end{figure}

  Figure \ref{figTrU4} shows $\lambda$-dependence of 
  $\frac{1}{3}\langle \mathrm{Tr} \ U_4 \rangle$ in generalized Landau gauge.
  For $\lambda \ne 0$, $\langle \mathrm{Tr} \ U_4 \rangle$ takes a non-zero real value, 
  and it approaches to zero continuously as $\lambda \rightarrow 0$.
  Here, it largely changes in the small region of $0 \le \lambda \lesssim 0.1$.
  The finiteness of $\langle \mathrm{Tr} \ U_4 \rangle$ is directly related to
  the infrared damping of the correlator $G_\lambda(R)$ 
  and the infrared form of the instantaneous potential $V_\lambda(R)$.

  Also for the 3Q system, the similar argument can be applied.
  For the large spatial separation of the three quarks, 
  $G_\lambda^{\rm 3Q}({\vec x}, {\vec y}, {\vec z})$ behaves asymptotically as
  \begin{eqnarray}
    G_\lambda^{\rm 3Q}({\vec x}, {\vec y}, {\vec z}) &\equiv& 
    \langle \varepsilon_{abc} \varepsilon_{a'b'c'} 
    U_4^{aa'}({\vec x}) U_4^{bb'}({\vec y}) U_4^{cc'}({\vec z}) \rangle \nonumber \\
    &\rightarrow& 
    \varepsilon_{abc} \varepsilon_{a'b'c'} 
    \langle U_4^{aa'} \rangle \langle U_4^{bb'} \rangle \langle U_4^{cc'} \rangle \nonumber \\
    &=& \frac{2}{9} \langle \mathrm{Tr} \ U_4 \rangle^3.
  \end{eqnarray}
  Therefore, for $\lambda \ne 0$, 
  $G_\lambda^{\rm 3Q}$ approaches to a non-zero finite constant, 
  and asymptotically gives a saturation of $V_\lambda^{\rm 3Q}$ in the deep-IR region.  

  \subsection{Asymptotic behavior of Polyakov-line correlator and finite-time potential}

  Next, we consider 
  $T$-length terminated Polyakov-line correlator
  $G_\lambda(R,T)$,
  which behaves asymptotically as
  \begin{eqnarray} 
    G_\lambda(R,T) &\equiv& 
    \langle \mathrm{Tr} L^\dagger(\vec{x},T)L(\vec{y},T)\rangle \nonumber \\
    &\rightarrow& \frac{1}{3} \langle \mathrm{Tr} \ L(T) \rangle^2
  \end{eqnarray} 
  for large separation of $R = |\vec{x}-\vec{y}|$.
  As well as the instantaneous potential,
  $\frac{1}{3} \langle \mathrm{Tr} \ L(T)\rangle^2$ 
  is found to give the lower bound of the correlator $G_\lambda(R,T)$,
  and the finiteness of $\langle \mathrm{Tr} \ L(T) \rangle$ 
  is responsible for the infrared saturation of the finite-time potential $V_\lambda(R,T)$.

\begin{figure}
    \centering
    \includegraphics[width=8cm,clip]{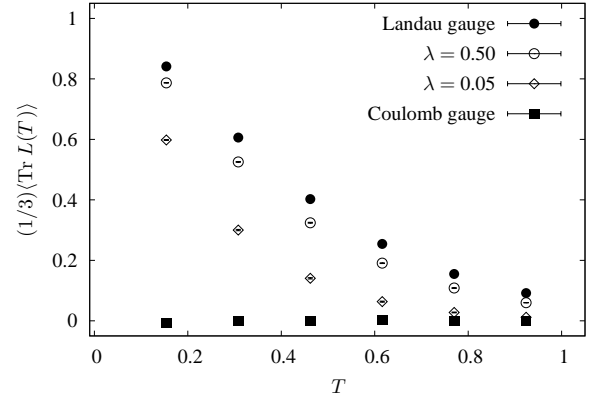} 
    \caption{ \label{figTermPol}
      $T$-dependence of 
      $\frac{1}{3}\langle \mathrm{Tr} \ L(T) \rangle$ in generalized Landau gauge.
      In the Coulomb gauge, $\langle \mathrm{Tr} \ L(T) \rangle$ is always zero.
      For $\lambda \neq 0$,
      $\langle \mathrm{Tr} \ L(T) \rangle$ is a decreasing function of $T$,
      and it converges to zero in large-$T$ limit.
    }
  \end{figure}

  Figure \ref{figTermPol} shows $T$-dependence of
  the terminated Polyakov-line $\frac{1}{3}\langle \mathrm{Tr} \ L(T) \rangle$ 
  in generalized Landau gauge
  for typical values of $\lambda$.
  In the Coulomb gauge ($\lambda$=0), $\langle \mathrm{Tr} \ L(T) \rangle$ is always zero
  as well as $\langle \mathrm{Tr} \ U_4 \rangle$, 
  which means that $V_\lambda(R=\infty,T) = -\frac{1}{T}{\rm ln} G_\lambda(R=\infty,T)=+\infty$ 
  for any values of $T$.
  Actually, finite-time potential $V_\lambda(R,T)$ always has a linear part 
  in the Coulomb gauge, as shown in Fig.\ref{figExtendedPot}.

  For $\lambda \neq 0$,
  $\langle \mathrm{Tr} \ L(T) \rangle$ is a decreasing function of $T$,
  and it converges to zero in large-$T$ limit.
  At $T = N_t$,
  the $T$-length terminated Polyakov-line $\langle \mathrm{Tr} \ L(T) \rangle$ 
   results in the Polyakov loop, and $\langle \mathrm{Tr} \ L(N_t) \rangle = 0$ in the confinement phase.
  Therefore, $\langle \mathrm{Tr} \ L(T)\rangle$ converges to zero as $T \rightarrow N_t$, 
  and then one finds  
  \begin{eqnarray} 
    G_\lambda(R=\infty,N_t)&=&0, \\
    V_\lambda(R=\infty,N_t)&=& -\frac{1}{T}{\rm ln} G_\lambda(\infty,N_t)=+\infty,
  \end{eqnarray} 
  which gives a confinement potential.
  From Fig.\ref{figTermPol}, this convergence is found to be fast for smaller $\lambda$-value, 
  and such a convergence is closely related to the growing of finite-time string tension $\sigma_\lambda(T)$.

  For the 3Q system, we can apply the similar argument.
  For the large spatial separation limit of the three quarks, 
  $G_\lambda^{\rm 3Q}({\vec x}, {\vec y}, {\vec z}, T)$ behaves asymptotically as
  \begin{eqnarray}
    G_\lambda^{\rm 3Q}({\vec x}, {\vec y}, {\vec z},T) &\equiv& 
    \langle \varepsilon_{abc} \varepsilon_{a'b'c'} 
    L^{aa'}({\vec x},T) L^{bb'}({\vec y},T) L^{cc'}({\vec z},T) \rangle \nonumber \\
    &\rightarrow& 
    \varepsilon_{abc} \varepsilon_{a'b'c'} 
    \langle L^{aa'}(T) \rangle \langle L^{bb'}(T) \rangle 
    \langle L^{cc'}(T) \rangle \nonumber \\
    &=& \frac{2}{9} \langle \mathrm{Tr} \ L(T) \rangle^3,
  \end{eqnarray}
  and non-zero $\langle \mathrm{Tr} \ L(T) \rangle$ leads to
  the infrared saturation of the finite-time 3Q potential 
  $V_\lambda^{\rm 3Q}(T)$, for $\lambda \ne 0$. 

  At $T = N_t$, because of the Polyakov-loop property of 
  $\langle \mathrm{Tr} \ L(N_t) \rangle = 0$ in the confinement phase, 
  one finds the confinement property for $T=N_t$ as
  \begin{eqnarray} 
    G_\lambda^{\rm 3Q}(N_t)&\rightarrow&0, \\
    V_\lambda^{\rm 3Q}(N_t)&=& -\frac{1}{T}{\rm ln} G_\lambda^{\rm 3Q}(N_t)\rightarrow+\infty,
  \end{eqnarray} 
  in the large limit of the three-quark spatial separation.

  \subsection{Gluon propagator 
    and instantaneous potential in the Landau gauge}

  In this subsection, we discuss the relation between the gluon propagator
  and the instantaneous potential in the Landau gauge.

  The gluon propagator is a two-point function
  of the gauge field $A_\mu(x)$, and is defined in Euclidean QCD as
  \begin{equation}
    D_{\mu\nu}(x,y) \equiv \langle \mathrm{Tr} A_\mu(x) A_\nu(y) \rangle,
  \end{equation}
  where the trivial color structure is dropped off by taking the trace.
  In the Landau gauge, we use the expression of $A_\mu$ in terms of $U_\mu$ as
  \begin{equation}
    A_4(x) = \frac{1}{2iag}[U_4(x)-U_4^\dagger(x)] + \mathcal{O}(a^2).
  \end{equation}
  Note that this expression is only justified 
  in the Landau gauge, or more generally in large-$\lambda$ gauges, 
  where the fluctuation of $A_4$ is highly suppressed.

  Then, the gluon propagator $D_{\mu\nu}$ is expressed using link-variables, e.g.,  
  \begin{eqnarray}
    \label{eqPropInU}
    &&a^2g^2 D_{44}(x,y)  
    = a^2 g^2 \langle \mathrm{Tr} A_4(x)A_4(y)\rangle \nonumber \\
    &\simeq& - \frac{1}{4}\langle \mathrm{Tr} 
    [U_4(x)-U_4^\dagger(x)][U_4(y)-U_4^\dagger(y)]\rangle \nonumber \\
    &=& \langle \mathrm{Tr} [U_4(x)U_4^\dagger(y)]\rangle \nonumber \\
    && \quad   - \frac{1}{4} 
         \langle \mathrm{Tr}[U_4(x)+U_4^\dagger(x)]
                            [U_4(y)+U_4^\dagger(y)]\rangle. 
  \end{eqnarray}
  The last term in Eq.(\ref{eqPropInU}) has only
  $O(a^4)$-order $(x-y)$-dependence,
  and actually it changes only a few $\%$ in the Landau gauge at $\beta$=5.8, 
  so that we here approximate this term as a constant $C$. Thus, Eq.(\ref{eqPropInU}) reduces  
  \begin{equation}
    \label{eqPropInst}
    a^2 g^2 D_{44}(x,y) \simeq \langle \mathrm{Tr} [U_4(x) U_4^\dagger(y)] \rangle
    - C,
  \end{equation}
  and we calculate the instantaneous potential $V_{\rm inst}(R)$ as 
  \begin{eqnarray}
    \label{eqVinstD44}
    V_{\rm inst}(R) &=& -\frac{1}{a} 
    \ln 
    \langle \mathrm{Tr} [U_4(\vec{x},a) U_4^\dagger(\vec{y},a)] \rangle \nonumber \\
    &=& - \frac{1}{a} \ln \left[
    C + a^2 g^2 D_{44}(R) \right] \nonumber \\
    &\simeq& - \frac{a g^2}{C} D_{44}(R) + \mathrm{const.}
  \end{eqnarray}
  In this way, the instantaneous potential $V_{\rm inst}(R)$ is
  expressed by using the $44$-component of the gluon propagator
  in the Landau and large-$\lambda$ gauges.

  In the previous work \cite{Iritani09}, 
  we have found that the Landau-gauge gluon propagator is well reproduced 
  by the four-dimensional Yukawa-function as 
  \begin{equation}
    D(r) \equiv D_{\mu\mu}(r) \propto \frac{1}{r}e^{-mr},
  \end{equation}
  with the Yukawa mass-parameter $m \simeq$ 0.6GeV,
  in the region of $r = 0.1 \sim 1$ fm.
  Apart from a prefactor from the tensor factor,
  $D_{44}(R)$ approximately behaves as the Yukawa-function,
  and therefore the instantaneous potential is expressed as
  \begin{equation}
    V_{\rm inst}(R) \simeq - \frac{A}{R}e^{-mR}
    + \mathrm{const.}
  \end{equation}
  in the Landau gauge.

  \begin{figure}
    \centering
    \includegraphics[width=8cm,clip]{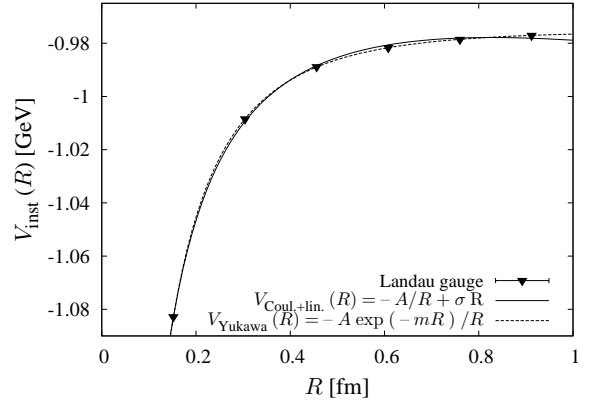}
    \caption{ \label{figInstLandau}
      Fit-result of the instantaneous potential $V_{\rm inst}(R)$ 
      in the Landau gauge
      using Coulomb plus linear form (solid line) 
      and Yukawa-function form (dashed line).
      Both forms well reproduce lattice QCD result.
      The best-fit parameter of the Yukawa mass is $m$ = 0.634(3)GeV,
      which coincides with the infrared effective gluon mass obtained from  
      the Landau-gauge gluon propagator \cite{Iritani09}.
    }
  \end{figure}

  Figure \ref{figInstLandau} shows the two fit-results of 
  instantaneous potential in the Landau gauge, 
  using the Coulomb plus linear form 
  $V_{\rm Coul.+lin.}(R) = -A/R + \sigma R$
  and the Yukawa-function form 
  $V_{\rm Yukawa}(R) = -A\exp(-mR)/R$.
  Both functions well reproduce the lattice QCD result.
  The best-fit Yukawa mass-parameter is $m$=0.634(3)GeV, 
  and this value coincides with the infrared effective gluon mass 
  obtained from the Landau-gauge gluon propagator \cite{Iritani09}.

  We note again that this relation is only valid in the Landau and
  large-$\lambda$ gauges, where
  the temporal link-variable $U_4$ can be expanded in terms of lattice spacing $a$,
  and the last-term in Eq.(\ref{eqPropInU}) is almost constant.

\section{Summary and Discussion}
  In this paper, aiming to grasp the gauge dependence of gluon properties,
  we have investigated generalized Landau gauge and applied it to 
  instantaneous interquark potential in SU(3) quenched lattice QCD at $\beta$=5.8.
  In the Coulomb gauge, the instantaneous potential is expressed by 
  the sum of Coulomb potential and linear potential 
  with 2-3 times larger string tension. 
  In contrast, the instantaneous potential has no linear part 
  in the Landau gauge. Thus, there is a large gap between these two gauges.
  Using generalized Landau gauge, we have found that the instantaneous 
  potential $V_\lambda(R)$ is connected continuously from the Landau gauge 
  towards the Coulomb gauge, and the linear part in $V_\lambda(R)$ 
  grows rapidly in the neighborhood of the Coulomb gauge.
  
  Since the slope $\sigma_\lambda$ of the instantaneous potential 
  $V_\lambda(R)$ grows continuously from 0 to 2-3$\sigma_{\rm phys}$, 
  there must exist some specific intermediate gauge 
  where the slope $\sigma_\lambda$ coincides with 
  the physical string tension $\sigma_{\rm phys}$. 
  From the lattice QCD calculation, the specific $\lambda$-parameter,
  $\lambda_C$, is estimated to be about $0.02$.

  We have also investigated $T$-length terminated Polyakov-line correlator, 
  and its corresponding finite-time potential $V_\lambda(R,T)$, 
  which is a generalization of the instantaneous potential $V_\lambda(R)$, 
  in generalized Landau gauge.

  In addition to the instantaneous/finite-time Q$\bar{\rm Q}$ potential,
  we have also analyzed the instantaneous/finite-time 3Q potential,
  which shows quantitatively similar behavior to the Q$\bar{\rm Q}$ case, 
  in terms of the gauge parameter $\lambda$ and the temporal length $T$.
  We thus consider that there is a universality of instantaneous/finite-time 
  potential for Q$\bar{\rm Q}$ systems and 3Q systems.

  Finally, we consider a possible gauge of QCD to describe 
  the quark potential model from the viewpoint of instantaneous potential. 
  The quark potential model is a successful nonrelativistic framework 
  with a potential instantaneously acting among quarks, 
  and describes many hadron properties in terms of quark degrees of freedom \cite{Isgur8586}. 
  In this model, there are no dynamical gluons, and gluonic effects 
  indirectly appear as the instantaneous interquark potential.
  In $\lambda_C$-gauge, the physical interquark potential
  $V_{\rm phys}(R)$ is approximately reproduced 
  by the instantaneous potential $V_{\lambda_C}(R)$, 
  so that, as an interesting possibility, 
  the $\lambda_C$-gauge may be a useful gauge 
  in considering the linkage from QCD to the quark potential model \cite{Iritani10}.
  In addition, since $\lambda_C \simeq 0.02 \ll 1$ 
  is a very small parameter in this framework, 
  it is interesting to apply  
  the perturbative technique in terms of $\lambda_C$ 
  for the calculation of the Faddeev-Popov determinant and so on.

\begin{acknowledgments}
  The authors thank Dr. H. Iida for his useful arguments.
  H.S. is supported in part by the Grant for Scientific
  Research [(C) No.19540287, Priority Areas ``New Hadrons'' (E01:21105006)] 
  from the Ministry of Education,
  Culture, Science and Technology (MEXT) of Japan.
  This work is supported by the Global COE Program,
  ``The Next Generation of Physics, Spun from Universality and Emergence".
  The lattice QCD calculations have been done on NEC-SX8 at
  Osaka University.
\end{acknowledgments}


\begin{thebibliography}{00}
\bibitem{Nambu66}
Y.~Nambu, {\it Proceedings of Preludes Theoretical Physics,
in honor of V.F.Weisskopf} (North-Holland, Amsterdam, 1966).

\bibitem{GWP73} 
D.J.~Gross and F.~Wilczek, Phys. Rev. Lett. {\bf 30}, 1343 (1973);
H.D.~Politzer, Phys. Rev. Lett. {\bf 30}, 1346 (1973).

\bibitem{KugoOjima}
T.~Kugo, and I.~Ojima, Suppl. Prog. Theor. Phys. {\bf 66}, 1 (1979); 
T.~Kugo, Proc. of Int. Symp. on ``BRS Symmetry on the Occasion
of Its 20th Anniversary'', 107, arXiv:hep-th/9511033.

\bibitem{Gribov}
V.~Gribov. Nucl. Phys. {\bf B139}, 1 (1978).

\bibitem{Zwanziger98}
D.~Zwanziger, Nucl.Phys. {\bf B518}, 237(1998).

\bibitem{Greensite03}
J.~Greensite, and S.~Olejn\'ik, Phys. Rev. D{\bf 67}, 094503 (2003);
J.~Greensite, Prog. Part. Nucl. Phys. {\bf 51}, 1 (2003).

\bibitem{Greensite04}
J.~Greensite, S.~Olejn\'ik, and D.~Zwanziger,
Phys. Rev. D {\bf 69}, 074506 (2004).

\bibitem{GreensiteThorn}
J.~Greensite and C.B.~Thorn, J. High Energy Phys. {\bf 02}, 014 (2002).

\bibitem{tHooft}
G.~'t~Hooft, Nucl. Phys. Proc. Suppl. {\bf 121}, 333 (2003); Nucl. Phys. {\bf A721}, 3 (2003). 

\bibitem{NambutHooftMandelstam}
Y.~Nambu, Phys. Rev. D{\bf 10}, 4262 (1974);
S.~Mandelstam, Phys. Rep. {\bf 23}, 245 (1976);
G.~'t Hooft, Nucl. Phys. {\bf B190}, 455(1981).

\bibitem{Bernard90}
C.~Bernard, D.~Murphy, A.~Soni, and K.~Yee, Nucl. Phys. B. (Proc. Suppl.) {\bf 17}, 593 (1990);
C.~Bernard, D.~Murphy, A.~Soni, Nucl. Phys. B (Proc. Suppl.) {\bf 20}, 410 (1991).

\bibitem{Iritani10}
T.~Iritani and H.~Suganuma, PoS({\bf LAT2010}), 277 (2010).

\bibitem{HigashimaMiransky}
V.A.~Miransky, ``Dynamical Symmetry Breaking in Quantum Field Theories''
(World Scientific, Singapore, 1993); 
K.~Higashijima, Prog. Theor. Phys. Suppl. {\bf 104}, 1 (1991).

\bibitem{Alkofer01}
R.~Alkofer and L.~von Smekal, Phys. Rept. {\bf 353}, 281 (2001) and its references.

\bibitem{Mandula99}
J. E. Mandula, Phys. Rept. {\bf 315},  273 (1999) and its references.

\bibitem{Iritani09}
T.~Iritani, H.~Suganuma, and H.~Iida,
Phys. Rev. {\bf D80}, 114505 (2009) and its references;
H.~Suganuma, T.~Iritani, A.~Yamamoto, and H.~Iida,
PoS({\bf QCD-TNT09}), 044 (2009); PoS({\bf LAT2010}), 289 (2010).

\bibitem{CGsource}
CP-PACS Collaboration (S. Aoki et al.). Phys. Rev. Lett. {\bf 84}, 238 (2000); 
Y.~Nemoto, N.~Nakajima, H.~Matsufuru, and H.~Suganuma, 
Phys. Rev. D{\bf 68}, 094505 (2003). 

\bibitem{ItzyksonZuber}
C.~Itzykson and J.~Zuber, ``Quantum Field Theory'',
(McGraw-Hill, New York, 1980).

\bibitem{Zwanziger03}
D.~Zwanziger, Phys. Rev. Lett. {\bf 90}, 102001 (2003).

\bibitem{STI}
H.~Suganuma, T.T.~Takahashi, and H.~Ichie,
``Color Confinement and Hadrons in Quantum Chromodynamics''
(World Scientific, Singapore, 2004), p. 249;
T.T.~Takahashi, H.~Suganuma, Y.~Nemoto, and H.~Matsufuru, 
Phys. Rev. D {\bf 65}, 114509 (2002);
T.T.~Takahashi, H.~Matsufuru, Y.~Nemoto, and H.~Suganuma, 
Phys. Rev. Lett. {\bf 86}, 18 (2001).

\bibitem{TS0304}
T.T.~Takahashi and H.~Suganuma, 
Phys. Rev. Lett. {\bf 90}, 182001 (2003);
Phys. Rev. D{\bf 70}, 074506 (2004);
F.~Okiharu, H.~Suganuma and T.T.~Takahashi, 
Phys. Rev. D{\bf 72}, 014505 (2005).

\bibitem{Nakamura06}
A.~Nakamura and T.~Saito,
Prog. Theor. Phys. {\bf 115}, 189 (2006).

\bibitem{Rothe}
H.J.~Rothe, ``Lattice Gauge Theories: An Introduction''
3rd ed. (World Scientific, Singapore, 2005).

\bibitem{Cucchieri07}
A.~Cucchieri, A.~Maas, and T.~Mendes,
Mod. Phys. Lett. {\bf A22}, 2429 (2007).

\bibitem{FuruiNakajima}
S.~Furui and H.~Nakajima,
Phys. Rev. D{\bf 69}, 074505 (2004).

\bibitem{Ichie}
H.~Ichie, V.~Bornyakov, T.~Streuer, and G.~Schierholz, Nucl. Phys. {\bf A721}, 899 (2003); 
V.G. Bornyakov, H. Ichie, Y. Mori, D. Pleiter, M.I. Polikarpov, 
G. Schierholz, T.~Streuer, H.~St\"uben, and T. Suzuki,
Phys. Rev. D{\bf 70}, 054506 (2004). 

\bibitem{BS04}
P.O.~Bowman and A.P.~Szczepaniak, Phys. Rev. D{\bf 70}, 016002 (2004). 

\bibitem{Isgur8586}
S.~Godfrey and N.~Isgur, Phys. Rev. D{\bf 32}, 189 (1985); 
S.~Capstick and N.~Isgur, Phys. Rev. D{\bf 34}, 2809 (1986). 

\end{thebibliography}
\end{document}